\documentclass[twocolumn,10pt]{revtex4}

\usepackage{amsmath}
\usepackage{amssymb}
\usepackage{amsfonts}
\usepackage{graphicx}
\usepackage{graphicx} % Include figure files
\usepackage{psfrag}					% remplacement du texte d'une figure ps par du texte latex
\usepackage{siunitx}					% mise en forme des unitŽs
\usepackage{color}					% gestion de diffŽrentes couleurs

\newcommand{\drawat}[3]{\makebox[0pt][l]{\raisebox{#2}{\hspace*{#1}#3}}}

\newcommand{\mt}{\tilde{m}}
\newcommand{\rt}{\tilde{r}}
\newcommand{\wo}{_{\mathrm{wo}}}
\newcommand{\w}{_{\mathrm{w}}}
\newcommand{\Awo}{$\mathrm{A_{wo}}$}
\newcommand{\Aw}{$\mathrm{A_{w}}$}
\newcommand{\Bwo}{$\mathrm{B_{wo}}$}
\newcommand{\Bw}{$\mathrm{B_{w}}$}
\newcommand{\tip}{\mathrm{tip}}

\definecolor{blue}{rgb}{0.3,0.3,1}
\newcommand{\correction}[1]{#1}

\renewcommand{\d}{\mathrm{d}}

\DeclareSIUnit \rtHz{$\sqrt{\mathrm{Hz}}$}

\begin{document}

%\preprint{}

\title{\correction{Functionalized {AFM} probes} for force spectroscopy: eigenmodes shape and stiffness calibration through thermal noise measurements} %Title of paper



\author{Justine Laurent, Audrey Steinberger and Ludovic Bellon}
\email[]{Ludovic.Bellon@ens-lyon.fr}
%\thanks{}
\affiliation{Universit\'e de Lyon, Laboratoire de Physique\\ \'Ecole Normale Sup\'erieure de Lyon, CNRS\\ 46 all\'ee d'Italie, FR 69007, Lyon, France}

\date{\today}

\begin{abstract}
The functionalization of an Atomic Force Microscope (AFM) cantilever with a colloidal bead is a widely used technique \correction{when} the geometry between the probe and the sample \correction{must} be controlled, \correction{particularly} in force spectroscopy. But some questions remain: how does a bead glued at the end of a cantilever influence its mechanical \correction{response} ? And more important for quantitative measurements, can we still determine the stiffness of the AFM probe with traditional techniques?

In this article, the influence of a colloidal mass loading on the eigenmodes shape and resonant frequency is investigated by measuring the thermal noise on rectangular AFM microcantilevers with and without a bead attached at their \correction{extremities}. The experiments are performed with a home-made ultra-sensitive AFM, based on differential interferometry. The focused beam from the interferometer probes the cantilever at different positions and the spatial shapes of the modes are determined up to the fifth resonance, without external excitation. The results clearly demonstrate that the first eigenmode almost doesn't change by mass loading. However the oscillation behavior of higher resonances present a marked difference: with a particle glued at its extremity, the nodes of the mode are displaced towards the free end of the cantilever. These results are compared to an analytical model taking into account the mass and the inertial moment of the load in an Euler-Bernoulli framework, where the normalization of the eigenmodes is explicitly worked out in order to allow a quantitative prediction of the thermal noise amplitude of each mode. A good agreement between the experimental results and the analytical model is demonstrated, allowing a clean calibration of the probe stiffness.

\end{abstract}


\maketitle 

\section{Introduction}

Atomic Force Microscopy (AFM) is currently used in a great variety of studies from various disciplines to measure small forces by measuring the deflection of a microcantilever~\cite{2005butt}. In biophysics\correction{,} for example, it has been applied to the unfolding of \correction{proteins}~\cite{1999Fisher,1999Carrion-Vazquez}, probing the structure of biological membranes~\cite{2009Frederix}, \correction{and} monitoring the mechanical response of living cells~\cite{2007Radmacher, 2008SBRANA}. In nanotechnology as well, micro-scale levers find applications in Micro-Electro-Mechanical Systems (MEMS) and other nanotechnological devices~\cite{Lavrik-2004}. In material, surface or nano sciences in general, AFM probes appear as a cornerstone for quantitative studies \correction{of nanoscale properties}~\cite{2008Bhushan-b}.

All those applications exploit the great accuracy in measuring the cantilever deflection offered by AFM and converting this measurement in units of force assuming the cantilever behaves like a spring with known stiffness. Manufacturers often specify the spring constant of their cantilevers in a wide range of values, mainly because of the great uncertainties in the dimensions, particularly the thickness, resulting from the fabrication process. To overcome this problem several techniques have been proposed to calibrate the cantilever spring constant~\cite{2005butt,cook,levy,1993hutter,sader-1999}. The reader is referred to the work of Burnham and co-workers~\cite{Burnham-2003} and the references therein for a comparative summary of the different techniques.

One of the first and still most commonly used calibration \correction{methods} is the so-called thermal calibration method based on the measurement of the vibration amplitude of the free end of a cantilever \correction{excited} by thermal noise~\cite{1993hutter}. The first peak of the thermal noise spectrum is related back to the spring constant of the cantilever modeled as an harmonic oscillator. In a more accurate model, Butt and Jaschke~\cite{butt} introduced a correction factor deduced from the Euler-Bernoulli description of the flexural dynamic of a free-clamped beam. In a previous work, we demonstrated that measuring this thermal noise for the first resonant modes of the cantilever provides an excellent benchmark to probe the mechanical response of the cantilever and compare it to a simple mechanical model~\cite{Paolino-JAP-2009}. 

In this article, we extend this method to the case of functionalized AFM cantilevers. Indeed, \correction{when} the geometry between the probe and the sample must be controlled, it is common to use a colloidal bead fixed at the free end of the lever. The radius of curvature of the ``tip'' is then controlled and stable, and offers a clean sphere-plane geometry to study interaction at nanometric distances. These modified probes (whether \correction{home-made or commercially available}) are commonly used in force spectroscopy, in particular for the measurement of the nanorheology of confined fluids~\cite{Maali2008,Maali2012} or of the Casimir interaction~\cite{Laurent-12-PRB}. How does this loaded mass influence the mechanical response of the AFM cantilever? Can the common techniques (thermal noise calibration in particular) to determine the stiffness still be used? 

In this work, we measure the thermal noise spectra of the cantilever deflection on its whole surface and compare the rms amplitudes obtained with and without a bead loaded at its free extremity. Furthermore, we compare the results to a simple mass model which modifies only the boundary conditions of the classical rectangular beam theory~\cite{Oguamanam}. A good agreement between experimental data and this analytical model will be demonstrated, showing that thermal noise calibration of the probe stiffness is still perfectly pertinent for such cantilevers. 

The paper is organized as follows. Section~\ref{section:theory} describes the theoretical approach, with a special emphasis on eigenmodes normalization to allow the prediction of thermal noise amplitude of each mode. Section~\ref{section:experiment} details the experimental results in the light of this model, for two cantilevers probing various mass ratios between the cantilever and the colloidal bead. \correction{A discussion and conclusions are} given in section~\ref{section:ccl}, with a specific focus on how our conclusions can be applied to the classic angular deflection measurement technique.

\section{Analytical description of thermal noise \label{section:theory}}

\subsection{Flexural eigenmodes of a clamped cantilever}

\begin{figure}[tb]
\begin{center}
\psfrag{experiment}{experiment}
\psfrag{model}{model}
\psfrag{x}{$x$}
\psfrag{x}{$x$}
\psfrag{y}{$y$}
\psfrag{z}{$z$}
\psfrag{L}{$L$}
\psfrag{W}{$W$}
\psfrag{T}{$T$}
\psfrag{z(x,t)}{$Z(x,t)$}
\psfrag{mb}{$m_b$}
\psfrag{rg}{$r_g$}
\includegraphics{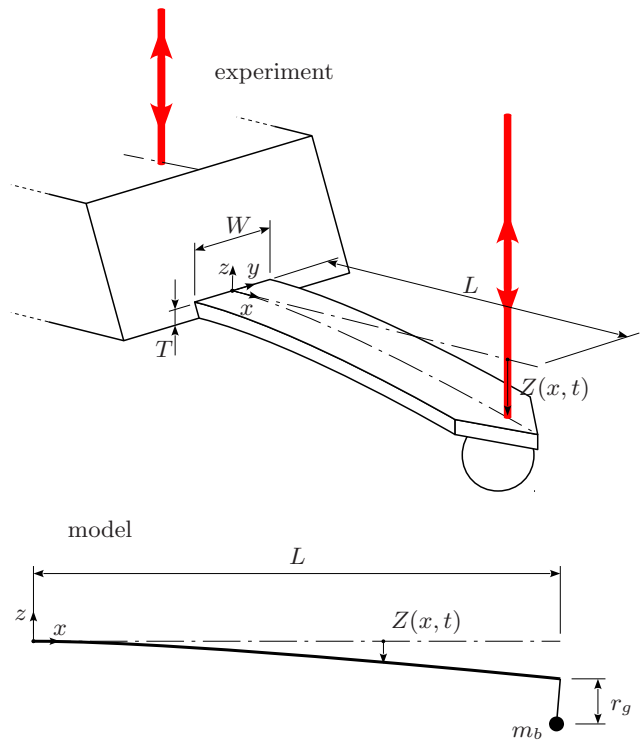}
\caption{Schematics of the experiment and analytical model. The cantilever (length $L$, width $W$, thickness $H$) is modeled in the Euler-Bernoulli framework by its neutral axis subject to a deflection $Z(x,t)$. The bead glued at its free end is modeled as a point mass $m_b$ located at the extremity. $r_g$ is the equivalent gyration radius of bead and accounts for the inertia in rotation of the bead at the cantilever end. In the experiment, the deflection can be measured at any position $x$ and $y$ (along and transverse to the cantilever axis) with a very low noise differential interferometer~\cite{Schonenberg-RSI-89,2002Bellon,Paolino-JAP-2009}, sensing the optical path difference between the two laser beams represented on the figure.}
\label{fig:cantilever}
\end{center}
\end{figure}

In the Euler-Bernoulli framework to describe the \correction{micrometer-sized} mechanical beam, we assume that the cantilever length $L$ is much larger than its width $W$, which itself is much larger than its thickness $T$ (see figure \ref{fig:cantilever}). The flexural modes of the cantilever are supposed to be only perpendicular to its length and uniform across its width. The deformations can thus be described by the deflection $Z(x,t)$, with $t$ the time, and $x$ the spatial coordinate along the beam normalized to its length $L$. The bead is supposed to be non deformable, and thus described as a rigid mass $m_b$ at the free end of the cantilever. It may be offset from the neutral axis, and carrying a non negligible inertia: these effects are taken into account with an inertial moment $m_b r_g^2$, $r_g$ being the equivalent gyration radius computed at the free end of the cantilever. For a sphere or radius $r$, $r_g$ \correction{is equal} to $\sqrt{7/5}\, r$. We neglect in our analysis any offset along the cantilever axis, and any coupling with torsion\correction{: precise gluing of the bead should make those effects negligible in first approximation. In order to reach a simple analytical solution, we also neglect the actual triangular shape of the cantilever at its free end: we will consider that the lever has an effective length smaller that its total length to approximate the real geometry. This last approximation will prove to be reasonable to describe our measurements.} Figure~\ref{fig:cantilever} sketches the experiment and the applied model. Following~\cite{Oguamanam}, we will include the effect of the bead in the boundary conditions of the cantilever dynamics.

The equation of motion for the cantilever, once the variables in time and space \correction{are} separated, can be written 
\begin{equation}
\frac{k}{3} \frac{\d^{4}z}{\d x^{4}}= m_c \omega^{2} z
\end{equation}
with $k$ the static stiffness of the cantilever, $m_{c}$ its mass, $Z(x,t)=z(x) e^{i\omega t}$ the deflection, and $\omega$ the pulsation. This equation can be rewritten as
\begin{equation} \label{ED}
z^{(4)}=\alpha^{4} z
\end{equation}
where $^{(n)}$ is the spatial derivative of order $n$, and $\alpha$ is given by the dispersion relation: 
\begin{equation} \label{DR}
\alpha^{4} =\frac{3 m_c \omega^{2}}{k}
\end{equation}
The generic solution to this equation is
\begin{equation} \label{deflexiongenerique0}
z(x)=a\cos(\alpha x) + b\sin(\alpha x) - c \cosh(\alpha x) - d \sinh (\alpha x)
\end{equation}
The boundary conditions in $x=0$ corresponds to a clamped end, implying $z(0)=0$ \correction{and} $z^{(1)}(0)=0$, hence $a=c$ and $b=d$. Defining $R=b/a$, \correction{expression} \ref{deflexiongenerique0} can thus be written 
\begin{equation} \label{deflexiongenerique}
z(x)=a\left(\cos(\alpha x) - \cosh(\alpha x) + R \left[\sin(\alpha x) - \sinh (\alpha x)\right]\right)
\end{equation}
\correction{For $x=1$, corresponding to the free end of the cantilever where the bead is glued,} the conditions on the force and torque are linked to the inertia in translation and rotation of the bead~\cite{Oguamanam}:
\begin{align}
 z^{(3)}(1) &=- \alpha^4 \tilde{m} z (1) \label{eq_e} \\
 z^{(2)}(1) &= \alpha^{4} \tilde{m} \tilde{r}^2 z^{(1)}(1) \label{eq_d}
\end{align}
where $\tilde{m}=m_b/m_c$ is the mass of the bead $m_b$ normalized to that of the cantilever $m_c$ , and $\tilde{r}=r_g/L$ is the gyration radius of the bead $r_g$ normalized to the cantilever length $L$.
Expressing those boundary conditions with expression \ref{deflexiongenerique} leads to
\begin{align} 
R &= \frac{\sin\alpha - \sinh \alpha + \alpha \tilde{m} (\cos\alpha - \cosh\alpha)}{\cos\alpha + \cosh\alpha - \alpha \tilde{m} (\sin\alpha - \sinh \alpha)} \\
&= - \frac{\cos\alpha + \cosh\alpha - \alpha^3 \tilde{m} \tilde{r}^2 (\sin\alpha + \sinh \alpha)}{\sin\alpha + \sinh \alpha + \alpha^3 \tilde{m} \tilde{r}^2 (\cos\alpha - \cosh\alpha)} 
\end{align}

\begin{figure}[tb]
\begin{center}
\include{modes_a}\includegraphics{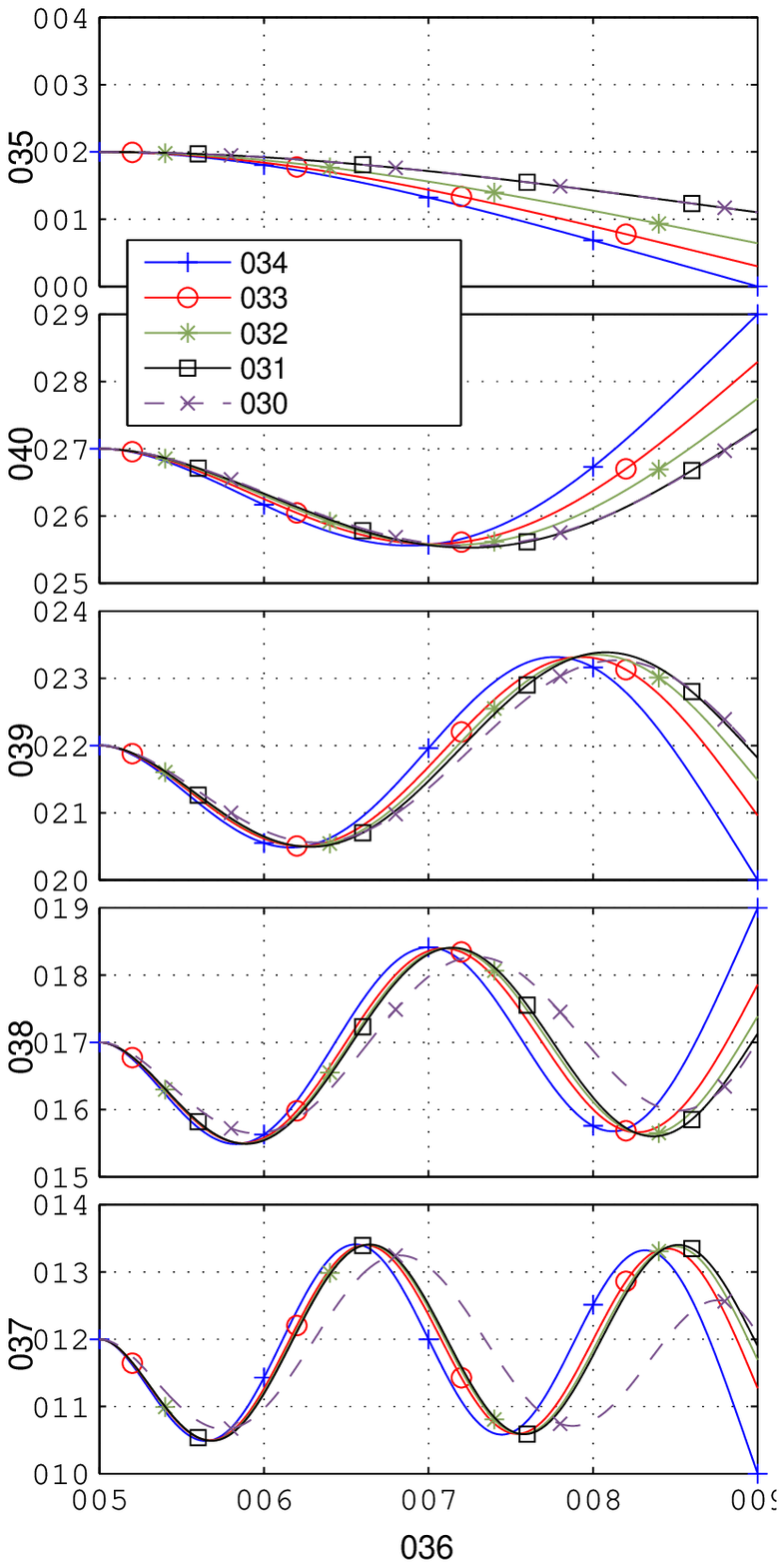}
\caption{\correction{(Colour online)} We plot in this figure the \correction{first 5} normal modes $\phi_{n}(x)$ for various $\tilde{m}$ and $\tilde{r}$. The amplitude of the first mode decreases and the nodes of the higher order modes get closer to the free end when \correction{either $\tilde{m}$ or $\tilde{r}$ increases}.}
\label{modes}
\end{center}
\end{figure}

The values of $\alpha$ allowing this equality are quantified, and correspond to the spatial eigenvalues $\alpha_n(\tilde{m},\tilde{r})$ of the resonant modes of the cantilever. They can be numerically computed. For $\tilde{m}=0$ (no bead), the last equation simplifies to the usual condition $1+\cos \alpha \cosh \alpha = 0$, leading to the common tabulated eigenvalues of a clamped-free Euler-Bernoulli mechanical beam. The $\alpha_n(\mt, \rt)$ values are reported for the \correction{first five} modes, for $0 \le \mt \le 2$ and $0 \le \rt \le 0.1$, in tables \correction{I toV in the supplementary data}. The corresponding shapes of the eigenmodes are plotted in figure \ref{modes}.

The length of the cantilever $L$, used in the normalization of $x$ and thus impacting the spatial eigenvalues $\alpha_n$, is sometimes experimentally ill-defined due to the triangular shape of the cantilever end. The direct comparison of the experimental values of $\alpha_n$ with the theoretical ones is thus hampered by this incertitude. However, their ratio is exempted from this bias, and can be used to check analytical predictions. In appendix \ref{appendix:plot-alpha}, we plot such ratios, useful to extract the values of $\mt$ and $\rt$ from the experimental observations.
%
%\begin{table}[htdp]
%\begin{center}
%\begin{tabular}{|l|r|r|r|r|r|}
%\hline
%$\tilde{m}$ & 0 & 0.1 & 0.3 & 1 & 1 \\
%\hline
%$\tilde{r}$ & 0 & 0 & 0 & 0 & 0.03 \\
%\hline
%$\alpha_{1} $ & 1.88 & 1.72 & 1.54 & 1.25 & 1.25 \\
%\hline
%$\alpha_{2} $ & 4.69 & 4.40 & 4.19 & 4.03 & 3.99 \\
%\hline
%$\alpha_{3} $ & 7.85 & 7.45 & 7.25 & 7.13 & 6.94 \\
%\hline
%$\alpha_{4} $ & 11.00 & 10.52 & 10.35 & 10.26 & 9.62 \\
%\hline
%$\alpha_{5} $ & 14.14 & 13.61 & 13.46 & 13.39 & 12.06 \\
%\hline\end{tabular}
%\end{center}
%\caption{Values of $\alpha_{n}$ for the first 5 eigenmodes of the cantilever for various $\tilde{m}$ and $\tilde{r}$.}
%\label{table:alpha_n}
%\end{table}%

\subsection{Orthogonality relations and normalization \label{section:orthogonality}}

The eigenmodes $z_n(x)$ are given by equation \ref{deflexiongenerique}, with a \correction{dependence} of the three parameters $a$, $R$ and $\alpha$ on mode number $n$ and on $\mt$ and $\rt$ (the \correction{dependence} in $\mt$ and $\rt$ will be implicit in our notations):
\begin{align} \label{zn}
z_{n}(x) & = a_n \zeta_n(x) \\
& = a_n \big(\cos(\alpha_n x)-\cosh(\alpha_n x) \nonumber \\
& \ \ \ \ \ \ \ \ + R_n \left[\sin(\alpha_n x)- \sinh(\alpha_n x) \right]\big)
\end{align}
However, if we only consider this expression, then the orthogonality between two modes $z_n$ and $z_m$ does not hold: it is easy to show that for $n\ne m$, 
\begin{equation}
\int_{0}^{1}z_{n}(x)z_{m}(x)\d x = - \tilde{m} z_{n}(1)z_{m}(1) - \tilde{m} \tilde{r}^2 z_{n}^{(1)}(1)z_{m}^{(1)}(1)
\end{equation}

\correction{We therefore define the normal modes by 
\begin{align} \label{defmodepropre}
\phi_{n}(x)=z_{n}(x) + & z_{n}(1) \sqrt{\tilde{m} \delta (x-1) } \nonumber \\
& + z_{n}^{(1)}(1) \rt \sqrt{\tilde{m} \delta (x-1+\epsilon)}
\end{align}
where $\delta (x-1)$ and $\delta (x-1+\epsilon)$ are the Dirac distributions centered on $x=1$ and $1-\epsilon$, $\epsilon$ being an arbitrarily small quantity only ensuring that there is no overlap between the $z_{n}(1)$ and $z_{n}^{(1)}(1)$ terms. With such definition, it is straightforward to prove the orthogonality of the $\phi_{n}$ basis. Moreover, as $\phi_{n}(x)=z_{n}(x)$ for any $x \in [0,1-\epsilon[$, $\phi_{n}(x)$ thus obeys the initial differential equation \ref{ED} and the boundary conditions in the limit $x=0$ and $x\rightarrow 1$ (in the limit $\epsilon \rightarrow 0$). Eventually, the normalization of $\phi_{n}$ is easily obtained by imposing 
\begin{align}
\int_0^1 \phi_n^2(x)dx & = 1 \\
a_{n}^{2}  \left(\int_{0}^{1}\zeta_{n}^{2}(x) \d x + \tilde{m} \zeta_{n}^{2}(1) + \tilde{m} \tilde{r}^2 \zeta_{n}'^{2}(1)\right)  & = 1 \label{norm1}
\end{align}
where $\zeta_n(x)$ is defined in equation \ref{zn}. This last equation thus imposes the values of $a_n$ to construct the orthonormal basis $\phi_{n}$. The result of this process is illustrated in figure~\ref{modes} for a few values of $\tilde{m}$ and $\rt$.}

Let us now give an energetic meaning to the Dirac term\correction{, in the case $\rt=0$}. We first compute the kinetic energy of a mode $\phi_n$ with amplitude $A$: $Z(x,t)=A\phi_{n}(x)\cos \omega_{n} t$. The speed of a mass element $m_c \d x$ is $-A\phi_{n}(x)\omega_{n}\sin \omega_{n} t$, the total kinetic energy is thus 
\begin{align}
E_{c} & = \int_{0}^{1} \frac{1}{2} \ m_c \d x \ A^{2} \phi_{n}^{2}(x) \omega_{n}^{2} \sin^{2} \omega_{n} t \\
& = \frac{1}{2} m_c \left(\int_{0}^{1} \phi_{n}^{2}(x) \d x\right) A^{2} \omega_{n}^{2} \sin^{2} \omega_{n} t \label{eq:Ec1}\\
& = \frac{1}{2} m_{c} A^{2} \omega_{n}^{2} \sin^{2} \omega_{n} t
\end{align}
This is the kinetic energy of \correction{a} harmonic oscillator of mass $m_c$, resonant pulsation $\omega_n$ and amplitude $A$. Notice that the amplitude of this oscillator is different from the deflection at the free end of the cantilever: $A_c=A \phi_{n} (x\rightarrow 1) = A z_n(1) = A a_{n} \zeta_{n} (1)$. We may also \correction{rewrite} the integral on $\phi_{n}^{2}(x)$ in equation \ref{eq:Ec1} using equation \ref{norm1}:
\begin{align}
E_{c} & = \frac{1}{2} m_{c} a_{n}^{2} \left(\int_{0}^{1} \zeta_{n}^{2}(x) \d x + \tilde{m} \zeta_{n}^{2}(1)\right) A^{2} \omega_{n}^{2} \sin^{2} \omega_{n} t \\
& = \int_{0}^{1} \frac{1}{2} m_c \d x \ A^{2} z_n^{2}(x) \omega_{n}^{2} \sin^{2} \omega_{n} t \nonumber \\
& \ \ \ \ \ \ \ \ \ \ \ \ + \frac{1}{2} m_b A_{c}^{2} \omega_{n}^{2} \sin^{2} \omega_{n} t \label{Ec0}
\end{align}
We easily identify here the sum of two terms: the kinetic energy of the mode $n$ of the cantilever itself subject to a sinusoidal motion with an amplitude $A _c$ at its free end, and the kinetic energy of a point mass $m_b$ subject to a sinusoidal motion with the same amplitude $A _c$. The additional Dirac term in $\phi_n$ thus takes into account the bead motion in the total energy of the equivalent harmonic oscillator. The amplitude of the latter is not equal to the amplitude at the free end of the cantilever (which is also the case without the added mass since $|\phi_{n}(x=1)|=2$ for $\mt=0$).

\begin{figure}[tb]
\begin{center}
\include{deflexionmodek0p1Npm}\includegraphics{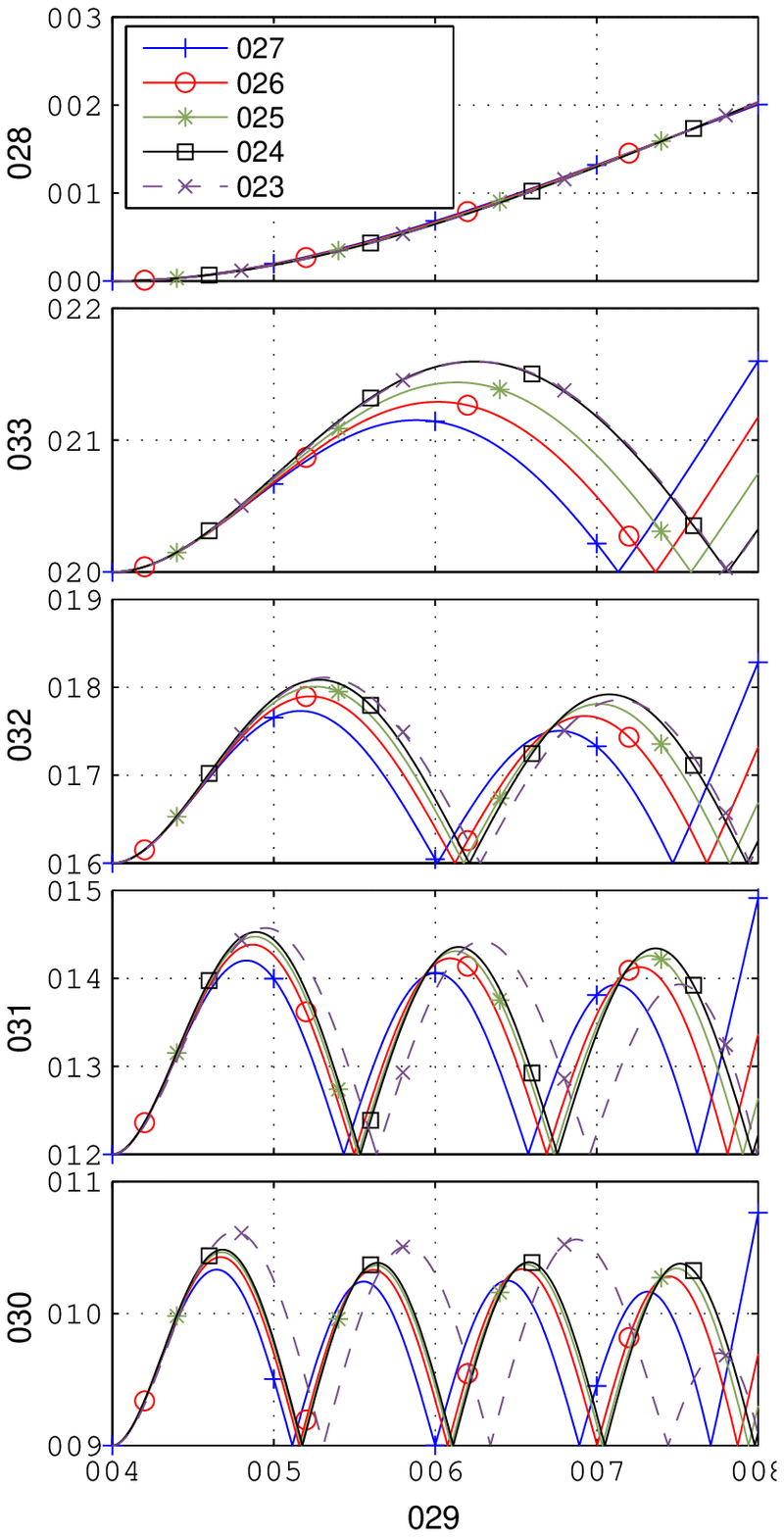}
\caption{\correction{(Colour online)} Thermal noise expected for a cantilever of static stiffness $k=\SI{0.1}{N/m}$ at $\SI{300}{K}$ for a few values of $\tilde{m}$ and $\tilde{r}$. The rms deflection of the first 5 modes is plotted as a function of the position along the cantilever $x$. The first mode is weakly impacted by the bead. When $\tilde{m}$ increases, the amplitude of the higher order modes vanish close to the free end, and the thermal energy is reported towards the antinodes, the amplitude of which raises.}
\label{deflexionmodek0p1Npm}
\end{center}
\end{figure}

\correction{In the case where $\tilde{r}\neq0$, we can extend this energetic approach to include the kinetic energy due to the rotation of the bead, and retrieve the normalization of the modes. A similar approach has been used by Oguamanam~\cite{Oguamanam} to ensure the orthonormalization of the normal modes in a more general framework including the coupling of flexural and torsional modes.}

\subsection{Thermal noise \correction{amplitude of resonant modes}}

\begin{figure}[tb]
\begin{center}
\include{energie}\includegraphics{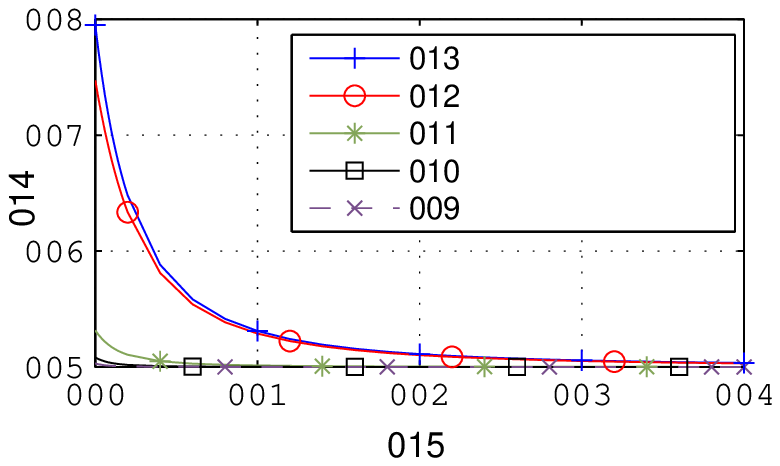}
\caption{\correction{(Colour online)} Fraction $\eta_{n}(x=1,\tilde{m},\tilde{r}=0)$ at the free end of the cantilever of the mean quadratic deflection driven by thermal noise for the first 5 modes as a function of the normalized bead mass $\tilde{m}$. The amplitude of the first mode (plotted as $1-\eta_{1}$ \correction{since $\eta_{1}\rightarrow 1$}) quickly overcomes the contribution of all the other modes, for which the inertia of the mass implies a node close to the free end of the cantilever.}
\label{energie}
\end{center}
\end{figure}

We compute the thermal noise of each resonant mode following~\cite{butt,Bellon-2008}: let us project the thermal noise driven deflection on the orthonormal basis $\phi_{n}(x)$:
\begin{equation}
Z(x,t)=\sum_{n=1}^{\infty} Z_{n}(t) \phi_{n}(x)
\end{equation}
Under the hypothesis of uncoupled modes, we have for each degree of freedom
\begin{equation}
\frac{1}{2}k_{n} \langle Z_{n}^{2}(t) \rangle = \frac{1}{2} k_{B} T
\end{equation}
where $k_{B}$ is Boltzmann's constant, $T$ the temperature of the cantilever, and $k_{n}$ the stiffness of the mode defined by:
\begin{equation} \label{eq:kn}
k_{n}= \frac{k}{3}\alpha_{n}^{4}=m_{c}\omega_{n}^{2}
\end{equation}
The mean quadratic deflection measured in $x$ should thus be 
\begin{align}
\langle Z^{2}(x,t) \rangle &=\sum_{n=1}^{\infty} \langle Z_{n} \rangle^{2}(t) |\phi_{n}(x)|^{2} \\
&= \frac{k_{B} T}{k} \sum_{n=1}^{\infty} 3 \frac{|\phi_{n}(x)|^{2}}{\alpha_{n}^{4}}= \frac{k_{B} T}{k} \sum_{n=1}^{\infty} \eta_{n} (x,\tilde{m},\tilde{r})
\end{align}
In figure~\ref{deflexionmodek0p1Npm}, we plot the expected rms thermal noise at $\SI{300}{K}$ along a cantilever of static stiffness $k=\SI{0.1}{N/m}$ for the  \correction{first five} eigenmodes, for a few values of $\tilde{m}$ and $\tilde{r}$. Note that the normalization of the $\phi_{n}$ basis is a crucial step to apply the energy equipartition theorem in a quantitative manner to this analysis. Our approach also allows us to estimate the repartition of energy between the different modes. For example, if we perform the measurement at the free end of the cantilever ($x=1$), the first mode \correction{accounts} for $\eta_{1}(x=1,\mt=0,\rt=0)=97\%$ of the total thermal fluctuations for $\tilde{m}=\tilde{r}=0$, and $\eta_{1}(x=1,\mt=1,\rt=0)=99.8\%$ for $\tilde{m}=1$ and $\tilde{r}=0$. In figure~\ref{energie}, we plot the contribution of each mode to the mean quadratic deflection measured at its extremity when $\tilde{m}$ changes (for $\tilde{r}=0$) : the larger the mass, the stronger is the contribution of the first mode.

\section{Experimental methodology and results \label{section:experiment}}

\subsection{Experiment description}

\begin{figure}
\begin{center}
\includegraphics[width=8.2cm]{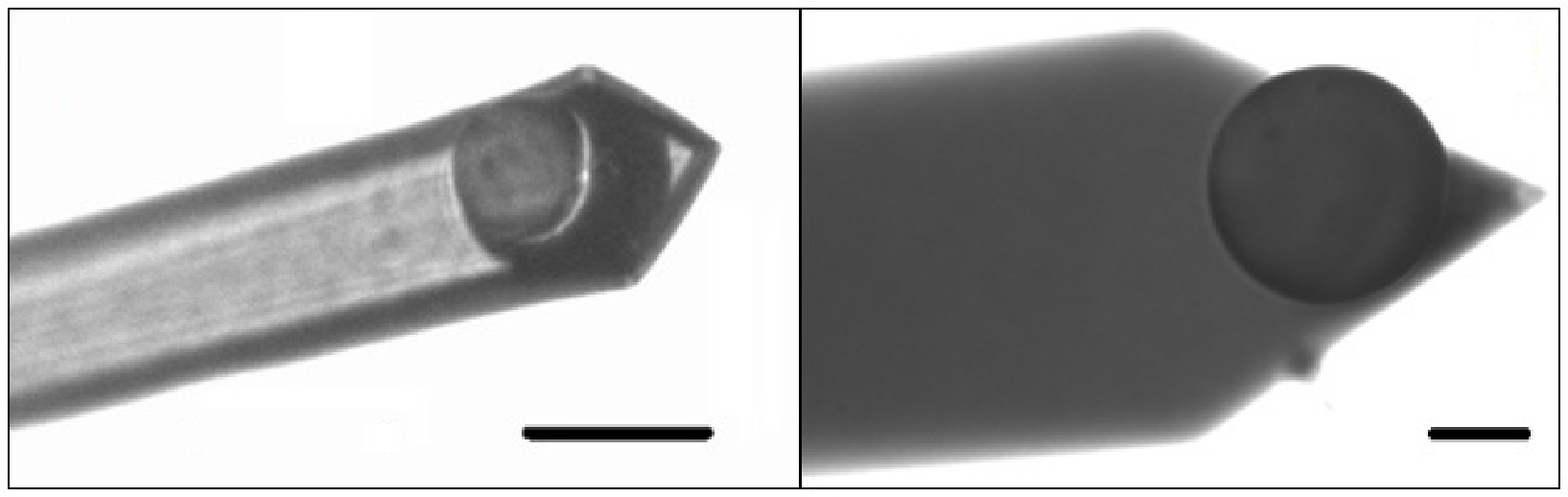}%
\drawat{-4.5cm}{2.2cm}{A}%
\drawat{-0.5cm}{2.2cm}{B}%
\drawat{-5.4cm}{0.55cm}{$\SI{20}{\micro m}$}%
\drawat{-0.96cm}{0.55cm}{$\SI{20}{\micro m}$}%
\caption{Pictures from an optical microscope of beads glued at the apex of cantilevers A and B.\label{fig1}} 
\end{center}
\end{figure}

\begin{table}[htdp]
\begin{center}
\begin{tabular}{|l|c|c|}
\hline
& Cantilever A & Cantilever B \\
\hline
Manufacturer & BudgetSensors & Nano-World \\
Reference & AIO, lever A & Arrow-TL8Au \\
\hline
Material & Silicon & Silicon \\
\hline
Tip height & $\SI{17\pm2}{\micro m}$ & tipless \\
\hline
Coating & none & Ti $\SI{5}{nm}$ + Au $\SI{30}{nm}$ \\
\hline
Resonant frequency & $\SI{15\pm5}{kHz}$ & $\SI{6}{kHz}$ \\
 & & $\SIrange[]{3}{14}{kHz}$\\
\hline
Force constant $k$ & $\SI{0.2}{N/m}$ & $\SI{0.03}{N/m}$\\
& $\SIrange{0.04}{0.7}{N/m}$ & $\SIrange{0.004}{0.54}{N/m}$ \\
\hline
Length $L$ & $\SI{500\pm10}{\micro m}$ & $\SI{500\pm5}{\micro m}$ \\
\hline
Width $W$ & $\SI{30\pm5}{\micro m} $ & $\SI{100\pm5}{\micro m}$ \\
\hline
Thickness $T$ & $\SI{2.7\pm1}{\micro m}$ & $\SI{1.0}{\micro m}$ \\
& & $\SIrange{0.5}{2.5}{\micro m}$ \\
\hline
Bead radius $r$ & $\SI{7.8\pm0.2}{\micro m}$ & $\SI{25.8\pm0.5}{\micro m}$ \\
\hline
\end{tabular}
\end{center}
\caption{Manufacturer specifications for cantilevers A and B. The last line corresponds to the glass bead that we glue at the free end of the cantilever, as illustrated in figure \ref{fig1}.}
\label{table:cantilevers}
\end{table}

Manufacturer specifications of our two \correction{samples} (A and B) are given in table~\ref{table:cantilevers}. Both present a ``rectangular'' geometry, close to the model used in our analytical approach. However, the triangular end (see figure~\ref{fig1}) departs from the model, and impedes a proper definition of their length $L$. We measure the deflection solely driven by thermal noise over the surface of these two different commercial cantilevers, first when they are still bare, then again after a glass bead has been glued at their free end \correction{with a two-part epoxy adhesive (Araldite). The bead of cantilever A is a fused borosilicate glass microsphere from Potters (Sphericel 110P8), whereas the bead B is a borosilicate sphere from Cospheric (BSGMS 45-53µm).} The radius $r$ of the bead is reported in table~\ref{table:cantilevers} for each sample.

As already mentioned, \correction{due to the manufacturing process,} the uncertainty in the thickness of the lever is large, resulting in a large uncertainty in the computation of its mass. The case is even worse for cantilever B, where the gold coating can change significantly the total mass due to the high density of gold. In addition, the quantity of glue cannot be measured precisely from the images of the cantilever. The geometric calculation of $\mt$ can thus only give a rough estimation of the actual value. We estimate $\mt_{A}\approx\num{0.1}$ (range $\numrange{0.04}{0.3}$) for cantilever A, and $\mt_{B}\approx1.2$ (range $\numrange{0.5}{2.6}$) for cantilever B. The geometric estimation of $\rt$ is less hampered by the uncertainty of the cantilever geometry, but still suffers from the uncontrolled repartition of the glue. We estimate $\rt_{A}=\num{0.02\pm0.01}$ for cantilever A, and $\rt_{B}=\num{0.06\pm0.02}$ for cantilever B. However, as demonstrated by \correction{Allen and collaborators}~\cite{Allen-2009}, the tip of the bare cantilever A itself may have a \correction{non-negligible} effect. We can geometrically estimate $\mt_{A\tip}\approx\num{0.04}$ and $\rt_{A\tip}\approx\num{0.02}$ for a $\SI{17}{\micro m}$ tall pyramidal tip.

The measurement is performed with a \correction{home-made} interferometric deflection sensor \cite{Paolino-JAP-2009}, inspired by the original design of Schonenberger \cite{Schonenberg-RSI-89} with a quadrature phase detection technique \cite{2002Bellon}: the interference between the reference laser beam reflecting on the chip of the cantilever and the sensing beam on the cantilever gives a direct measurement of the deflection with very high accuracy (see figure~\ref{fig:cantilever}). This technique offers a very low intrinsic noise (down to $\SI{E-14}{m/\rtHz}$). It is intrinsically calibrated as it measures directly the deflection against the wavelength of the laser beam, contrary to the standard optical lever technique that actually measures an angular deflection. \correction{Lastly}, the focused beam size resolution is tuned to as small as $\SI{10}{\micro m}$ to ensure a good spatial resolution.

 \correction{We apply the methodology of Paolino and coworkers~\cite{Paolino-JAP-2009}:} at every position $x$ and $y$ on a $\SI{5}{\micro m} \times \SI{5}{\micro m}$ grid, we measure the deflection $z(x,y,t)$ produced by the sole thermal excitation of the cantilever and we evaluate the power spectrum density (PSD) $S_z(x,y,f)$ on a $\SI{20}{s}$ signal sampled at $\SI{2}{MHz}$. For a quantitative characterization of the shape of the modes, the mean squared amplitude of each resonance $\left\langle A_n^2 (x,y) \right\rangle$ is determined as a function of positions $x$ and $y$ by integrating the PSD in a convenient frequency interval $2\Delta f$ around each resonance frequency $f_n$:
\begin{equation}
 \left\langle A_n^2 (x,y) \right\rangle = \int_{f_n-\Delta f}^{f_n+\Delta f} S_z^2(x,y,f) df.
 \label{eq1}
\end{equation}
This quantity is computed directly from the experimental spectra, without any fitting process. We take care to subtract the background noise contribution of the interferometer, and we also compensate for finite integration range in frequency~\cite{Paolino-JAP-2009}. 

\begin{figure*}
\begin{center}
\includegraphics[width=16cm]{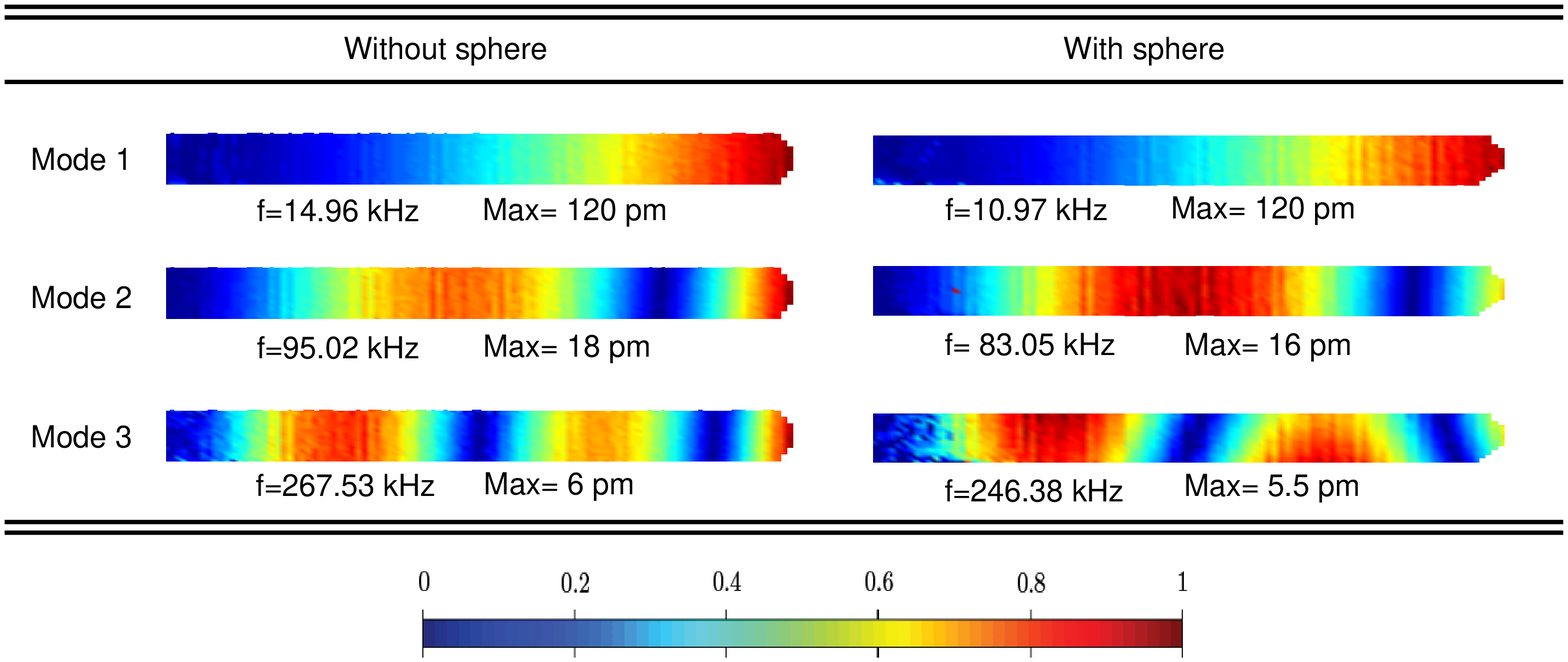}
\caption{\correction{(Colour online)} Maps of thermal noise of cantilever A. The rms amplitude of the first 3 flexural modes, with (right) and without (left) a bead glued at the free cantilever end, are plotted using a color coded amplitude map (color bar at the bottom of the figure). The resonant frequency and full scale amplitude are given below each map. A weak component in torsion can be \correction{seen clearly} or the third mode with the bead, hinting at a slightly off axis gluing of the colloidal particle.\label{fig2}} 
\end{center}
\end{figure*}

The complete set of results for cantilever A is reported in figure~\ref{fig2}, where the rms amplitude $\sqrt{\left\langle A_n^2 (x,y) \right\rangle}$ is represented with a color coded scale. The first three vibration modes can be clearly seen with their respective number of nodes. A weak component in torsion can be seen for the third mode with the bead, certainly because it has not been glued perfectly on the axis. However, in the following, we will neglect this effect and focus on the flexural modes along the $x$ axis. Therefore, at each position $x$, the median $\sqrt{\left\langle A_n^2 (x) \right\rangle}$ along the $y$ axis of $\sqrt{\left\langle A_n^2 (x,y) \right\rangle}$ is calculated. Due to the higher reflectivity of  \correction{gold-coated} cantilever B, the background noise is lower and we accurately measure the thermal noise up to the fifth resonant mode for this last sample.

\subsection{Results and discussion}

\subsubsection{Resonant frequency ratios\label{section:freqr}}

In a first attempt to estimate the added mass parameters $\mt$ and $\rt$, which must be known for proper normalization of the normal modes (value of $a_{n}$), let us first focus on the ratio between the resonant frequency $f_{n}$ of the successive modes. Indeed, equation \ref{eq:kn} translates into:
\begin{equation}
\frac{\alpha_{n}(\mt,\rt)}{\alpha_{1}(\mt,\rt)}=\sqrt{\frac{f_{n}}{f_{1}}}
\end{equation}
The resonant frequencies of each mode are easily found by a simple harmonic oscillator fit of each resonant peak in the thermal noise spectrum. We report in table \ref{table:freq} the square root of the frequencies of the first 3 modes of cantilever A and the first 5 modes of cantilever B, normalized to the frequency of the mode 1. We can then compare those measurements to the output of the analytical model (figure \ref{alphar}), and try to estimate the values of $\mt$ and $\rt$ for our samples. 

For cantilever A without the bead, the presence of the tip can be detected on the measured modes, and \correction{leads} to the estimation $\mt_{A\tip}=\num{0.057\pm0.010}$ and $\rt_{A\tip}=\num{0.03\pm0.02}$ (the standard deviations correspond to a $\SI{0.2}{\%}$ max distance between the analytical model and measured frequency ratio on the 3 modes), close to the values expected from the geometrical analysis. Adding the bead  \correction{changes} those values to $\mt_{A}=\num{0.336\pm0.008}$ and $\rt_{A}=\num{0.029\pm0.005}$ (same criterium for error bars).

For bare cantilever B, the ratios of frequencies between modes cannot be explained by our model : the triangular shape of the end of the cantilever alters the results corresponding to a rectangular one. If\correction{, however,} we look at the data for the loaded cantilever, we have a reasonable agreement for $\mt_{B}=\num{1.18\pm0.09}$ and $\rt_{B}=\num{0.070\pm0.006}$ (the  standard deviations correspond to a $\SI{2}{\%}$ max distance between the analytical model and measured frequency ratio on the 5 modes), again in line with the geometrical analysis.

\subsubsection{Resonant frequency shifts - a naive attempt\label{section:freq1}}

\begin{table*}[htdp]
\begin{center}
\begin{tabular}{|l||c|c|c||c|c|c|c|c|}
\hline
Bare cantilever & \multicolumn{3}{c||}{A: $\mt_{A\tip}=0.057$, $\rt_{A\tip}= 0.03$} & \multicolumn{5}{c|}{B: $\mt_{B\tip}=0$, $\rt_{B\tip}= 0$} \\
\hline
Mode number $n$ & \makebox[1.5cm][c]{1} & \makebox[1.5cm][c]{2} & \makebox[1.5cm][c]{3} & \makebox[1.5cm][c]{1} & \makebox[1.5cm][c]{2} & \makebox[1.5cm][c]{3} & \makebox[1.5cm][c]{4} & \makebox[1.5cm][c]{5}  \\
\hline
$f_{n}^{\wo}$ ($\SI{}{Hz}$) & \num{14956} & \num{95017} & \num{267530} & \num{7195} & \num{42975} & \num{117220} & \num{224500} & \num{363370} \\
\hline
$\sqrt{f_{n}^{\wo}/f_{1}^{\wo}}$ & 1.000 & 2.521 & 4.229 & 1.000 & 2.444 & 4.036 & 5.586 & 7.107 \\
\hline
$\alpha_{n}(\mt_{\tip},\rt_{\tip})/\alpha_{1}(\mt_{\tip},\rt_{\tip})$ & 1.000 & 2.521 & 4.236 & 1.000 & 2.503 & 4.189 & 5.864 & 7.539 \\
\hfill \emph{\small disagreement} & 0.0\% & -0.0\% & -0.1\% & 0.0\% & -2.4\% & -3.6\% & -4.7\% & -5.7\% \\
\hline
\hline
Loaded cantilever & \multicolumn{3}{c||}{A: $\mt_{A}=0.35$, $\rt_{A}= 0.03$} & \multicolumn{5}{c|}{B: $\mt_{B}=1.18$, $\rt_{B}= 0.06$} \\
\hline
Mode number $n$ &1 & 2 & 3 &1 & 2 & 3 & 4 & 5 \\
\hline
$f_{n}^{\w}$ ($\SI{}{Hz}$) & \num{10974} & \num{83050} & \num{246380} & \num{3101} & \num{29319} & \num{79023} & \num{152400} & \num{270980} \\
\hline
$\sqrt{f_{n}^{\w}/f_{1}^{\w}}$ & 1.000 & 2.751 & 4.738 & 1.000 & 3.075 & 5.048 & 7.010 & 9.348 \\
\hline
$\alpha_{n}(\mt,\rt)/\alpha_{1}(\mt,\rt)$ & 1.000 & 2.762 & 4.762 & 1.000 & 3.195 & 5.190 & 7.040 & 9.347 \\
\hfill \emph{\small disagreement} & 0.0\% & -0.4\% & -0.5\% & 0.0\% & -3.8\% & -2.7\% & -0.4\% & 0.0\% \\
\hline
\hline
Frequency shift upon loading & \multicolumn{3}{c||}{cantilever A} & \multicolumn{5}{c|}{cantilever B} \\
\hline
Mode number $n$ &1 & 2 & 3 &1 & 2 & 3 & 4 & 5 \\
\hline
$(L^{\w}/L^{\wo})\sqrt{f_{n}^{\w}/f_{n}^{\wo}}$ & 0.839 & 0.916 & 0.940 & 0.643 & 0.809 & 0.805 & 0.807 & 0.846 \\
\hline
$\alpha_{n}(\mt,\rt)/\alpha_{n}(\mt_{\tip},\rt_{\tip})$ & 0.844 & 0.924 & 0.948 & 0.642 & 0.820 & 0.796 & 0.771 & 0.796 \\
\hfill \emph{\small disagreement} & -0.5\% & -0.9\% & -0.8\% & 0.2\% & -1.2\% & 1.1\% & 4.7\% & 6.3\% \\
\hline
\end{tabular}
\end{center}
\caption{Frequency ratio between modes and frequency shift upon gluing of the bead. The square root of these ratios should be equal to the ratio of the corresponding spatial eigenvalues $\alpha_{n}(\mt,\rt)$ (with a correcting factor $L^{\w}/L^{\wo}$ for the frequency shift, that we suppose equal to $0.98$ here). Using estimated values of $\mt$ and $\rt$ for each measurement set (bare and loaded cantilevers A and B), we get a good overall agreement for every mode, especially for sample A. The \correction{model} reaches its limitation for the higher modes of cantilever B, whose triangular shaped end is not taken into account.}
\label{table:freq}
\end{table*}

Another way to determine the $\mt$ and $\rt$ parameters is to \correction{analyze} the resonant frequency shifts due to the loading, in an approach similar to the Cleveland method~\cite{Cleveland-1993}. In a naive attempt, we suppose that the process of gluing a bead to the cantilever free-end should have a limited effect on its stiffness and proper mass, thus we can relate the frequencies with ($f_n^\mathrm{w}$) and without ($f_n^\mathrm{wo}$) the mass to the spatial eigenvalues with ($\alpha_n(\mt,\rt)$) and without ($\alpha_n(0,0)$) through equation \ref{eq:kn}:
\begin{equation} \label{eq:fnwwo-naive}
\frac{\alpha_{n}(\mt,\rt)}{\alpha_{n}(0,0)}=\sqrt{\frac{f_{n}^\mathrm{w}}{f_{n}^\mathrm{wo}}}
\end{equation}
The resonant frequencies of each mode are easily found by a simple harmonic oscillator fit of each resonant peak of the thermal noise spectrum. We can then compare those measurements to the output of the analytical model, and try to estimate the values of $\mt$ and $\rt$ for our samples.

As shown in figure \ref{alphan}, the first mode is almost independent \correction{of} the value of $\rt$, and should thus reliably be used to measure $\mt$. This measurement is equivalent to the method of the added mass proposed by Cleveland~\cite{Cleveland-1993}. We find\correction{, with this protocol,} $\mt=\num{0.21\pm0.005}$ for cantilever A, and $\mt=\num{1.065\pm0.01}$ for cantilever B. 

This estimation of $\mt$ is hardly compatible with the expectation of the higher order modes for cantilever A, which would rather be $\tilde{m}=\num{0.10\pm0.01}$ (mode 2) or $\tilde{m}=\num{0.055\pm0.01}$ (mode 3), even considering the effect of $\rt$. As for cantilever B, the estimation could be compatible with higher order modes, but for both \correction{cantilevers} the value of $\mt$ is clearly underestimated with respect to the previous measurement through the frequency ratios between successive modes of the loaded cantilever.

\subsubsection{Resonant frequency shifts - refined analysis\label{section:freq2}}

Two naive hypotheses are responsible for the short-comings of the previous analysis of the frequency shifts upon loading: it first relies on the assumption that the effective length of the loaded cantilever is unchanged. It then assumes that the behavior of the cantilever without the load is that \correction{of} an ideal bare rectangular cantilever 

Let us first consider the effect of a possible modification of the effective length $L$ of the cantilever upon gluing a bead close to its free end. Indeed, this process may rigidify the end portion of the cantilever, thus shortening its effective length by the rigid part. Alternatively, the inertia of a large mass not fixed exactly at the free end, by bringing the nodes of the higher order modes closer to the position of the bead than to the free end, can also lead to a shortening of the effective length of the loaded cantilever. In order to evidence the dependence of the eigenvalue $\alpha_{n}$ in the cantilever's length $L$, equation~\ref{DR} can be expressed as:
\begin{equation} \label{DR_L}
\alpha_n^{4} =\frac{3 \mu L^4 \omega_n^{2}}{EI}
\end{equation}
where $E$ is the cantilever's Young modulus, $I$ its second moment of inertia, and $\mu$ its mass per unit length. As $E$, $I$ and $\mu$ do not depend \correction{either} on the cantilever's length or on the gluing of a bead, equation \ref{eq:fnwwo-naive} should in fact \correction{be written}
\begin{equation} \label{eq:fnwwo}
\frac{\alpha_{n}^\mathrm{w}}{\alpha_{n}^\mathrm{wo}}=\frac{L^\mathrm{w}}{L^\mathrm{wo}}\sqrt{\frac{f_{n}^\mathrm{w}}{f_{n}^\mathrm{wo}}}
\end{equation}
where the superscript $^\mathrm{w}$ (respectively $^\mathrm{wo}$) designates a quantity with (respectively without) the load. Assuming $\alpha_{n}^\mathrm{wo}$ is known, the parameters $\mt$ and $\rt$ for the loaded cantilever can be deduced by comparing the $\alpha_{n}^\mathrm{w}$ values obtained from equation~\ref{eq:fnwwo} to the tabulated values given in appendix~\ref{appendix:plot-alpha}\footnote{In order to retrieve the variation of the resonant frequencies with the fixation distance of a given bead to the free end of a cantilever, or to determine the added mass $m_b$ and the gyration radius $r_g$ from the $\mt$ and $\rt$ values, one must remember than the values of $\mt$ and $\rt$ determined from the resonant frequencies ratios or shifts also depend on the effective length of the loaded cantilever.}. Let us stress that if the resonant frequencies are measured with a very good accuracy, the effective length ratio is not known a priori. A small error in the effective length ratio leads only to the same relative error on the value of $\alpha_{n}^\mathrm{w}$. But since $\alpha_{n}(\mt,\rt)$ varies only slowly with $\mt$ (see appendix~\ref{appendix:plot-alpha}), it can result in a much larger error on $\mt$ and thus on the normalization of the \correction{normal modes} (parameter $a_n$). \correction{In section \ref{subsubsection:stiffness}}, we estimate that the gluing of a bead reduces the effective length of cantilevers A and B by $\SI{10}{\micro m}$, resulting in a $2\%$ shortening.  We thus use this $2\%$ correction in table \ref{table:freq} to compare the relative frequency shifts to the \correction{spatial} eigenvalues ratios.

When \correction{analyzing} the frequency ratios with and without a bead, it is also very important to use the right $\alpha_{n}^\mathrm{wo}$ value for the cantilever without the bead. We have seen with the naive approach that ignoring this initial loading for cantilever A leads to inconsistent values between modes, underestimating the true loading. However, if $\alpha_{n}(\mt_{A\tip},\rt_{A\tip})$ (with $\mt_{A\tip}=\num{0.057}$ and $\rt_{A\tip}=\num{0.03}$ as determined in the first section) is used as the unloaded reference value instead of $\alpha_{n}(0,0)$, one obtains $\mt_{A}=\num{0.37\pm0.04}$ and $\rt_{A}=\num{0.031\pm0.018}$ with the first three modes, in much better agreement with the previous estimation. 

The behavior of the tipless cantilever B also deviates from the one of a bare rectangular cantilever because of its triangular end. However, since the frequency ratios between the modes of the unloaded cantilever B do not yield any consistent set of $\mt$ and $\rt$ values (see figure~\ref{alphar}), we choose to take $\mt_{B\tip}=\rt_{B\tip}=0$ for the bare cantilever B. We thus compare the square root of the frequency ratio (corrected by the length ratio) with the $\alpha_{n}(\mt,\rt)/\alpha_{n}(0,0)$ ratio displayed on figure~\ref{alphan} for the \correction{first five} modes of cantilever B. As shown in figure \ref{alphan}, the first mode is almost independent \correction{of} the value of $\rt$, and can thus be used to measure $\mt$ alone; we obtain $\mt_{B}=\num{1.17\pm0.04}$ for mode 1. The higher order modes are compatible with the estimation of $\mt_{B}=1.17$ and can be used to guess the value of $\rt_{B}$. Using figure \ref{alphan}, we measure $\tilde{r}_{B} = 0.068$ (mode 2), $\tilde{r}_{B} = 0.057$ (mode 3), $\tilde{r}_{B} = 0.046$ (mode 4) \correction{and} $\tilde{r} _{B}= 0.030$ (mode 5). The dispersion of results is quite large for $\tilde{r}_{B}$, and points to the limitations of the model with respect to the actual cantilever shape. A simultaneous least \correction{squares} minimization of the distance between the analytical model and measured frequency shifts on the 5 modes leads to : $\mt_{B}=\num{1.19\pm0.13}$ and $\rt_{B}=\num{0.051\pm0.006}$ (standard deviation corresponding to a $\SI{3.5}{\%}$ max distance).

As a summary, we have two ways to estimate the normalized added mass and equivalent gyration radius from the measurement of the resonant frequencies of the loaded and unloaded cantilever: the frequency ratio between modes in one measurement, and the frequency shifts due to the addition of the bead. Provided the initial tip and effective length shortening are taken into account, both methods agree reasonably, though the \correction{dispersion} on $\rt$ is quite large. In the following we will retain the values :
\begin{itemize}
\item Unloaded cantilever A:  $\mt_{A\tip}=0.057$, $\tilde{r}_{A\tip}= 0.03$.
\item Loaded cantilever A:  $\mt_{A}=0.35$, $\tilde{r}_{A}= 0.03$.
\item Unloaded cantilever B:  $\mt_{B\tip}=0$, $\tilde{r}_{B\tip}= 0$.
\item Loaded cantilever B:  $\mt_{B}=1.18$, $\tilde{r}_{B}= 0.06$.
\end{itemize}

Eventually, using those estimations of $\mt$ and $\rt$, we can compute the values of the eigenvalues $\alpha_n(\mt,\rt)$ from the model, and compare them with the frequency shifts due to the bead. As one can read in table \ref{table:freq}, the agreement is quite good, with an overall agreement better than 1\% for cantilever A, and 3\% for cantilever B (except for the highest order modes of cantilever B, where the limitations of the model appear more severely).

\subsubsection{Spatial modes shapes}

\begin{figure}
\begin{center}
\include{cantileverA}\includegraphics{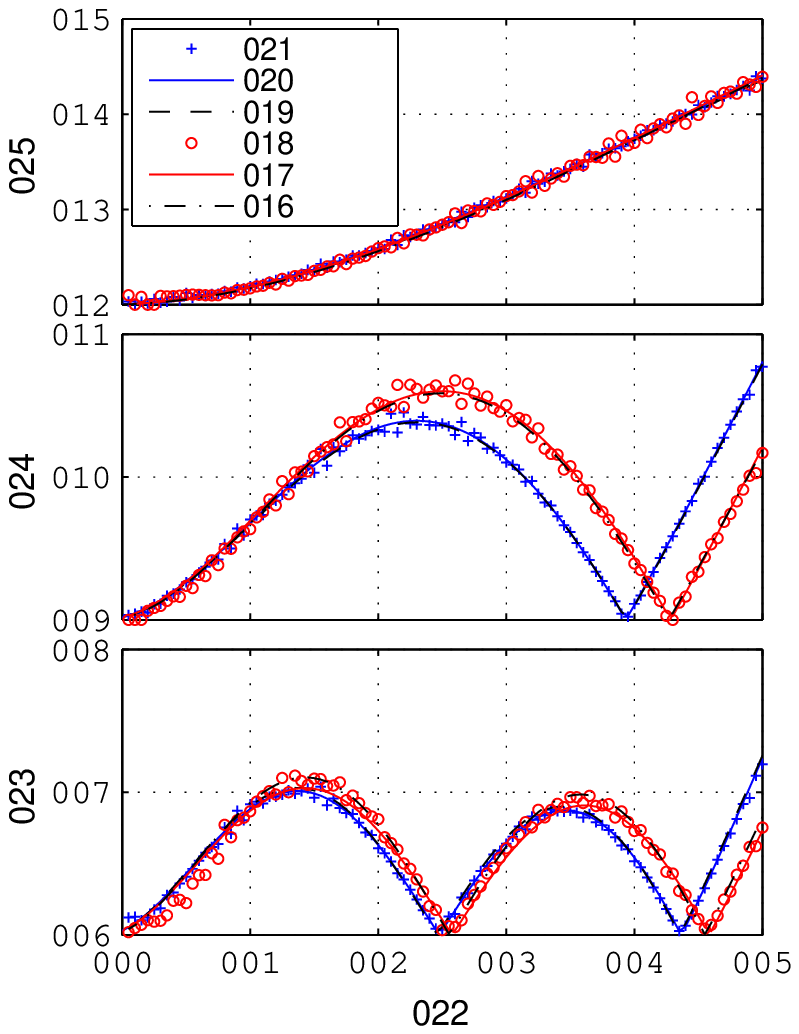}
\caption{\correction{(Colour online)} Amplitude of thermal noise for the first 3 flexural modes along the cantilever A with and without a sphere (in red and blue respectively). The markers represent the data while the lines exhibit the fits : independent fits of each mode in plain line, and simultaneous fit of all modes in dashed line. The agreement is excellent for the bare and the loaded cantilever.\label{fig3}} 
\end{center}
\end{figure}

\begin{figure}
\begin{center}
\include{cantileverB}\includegraphics{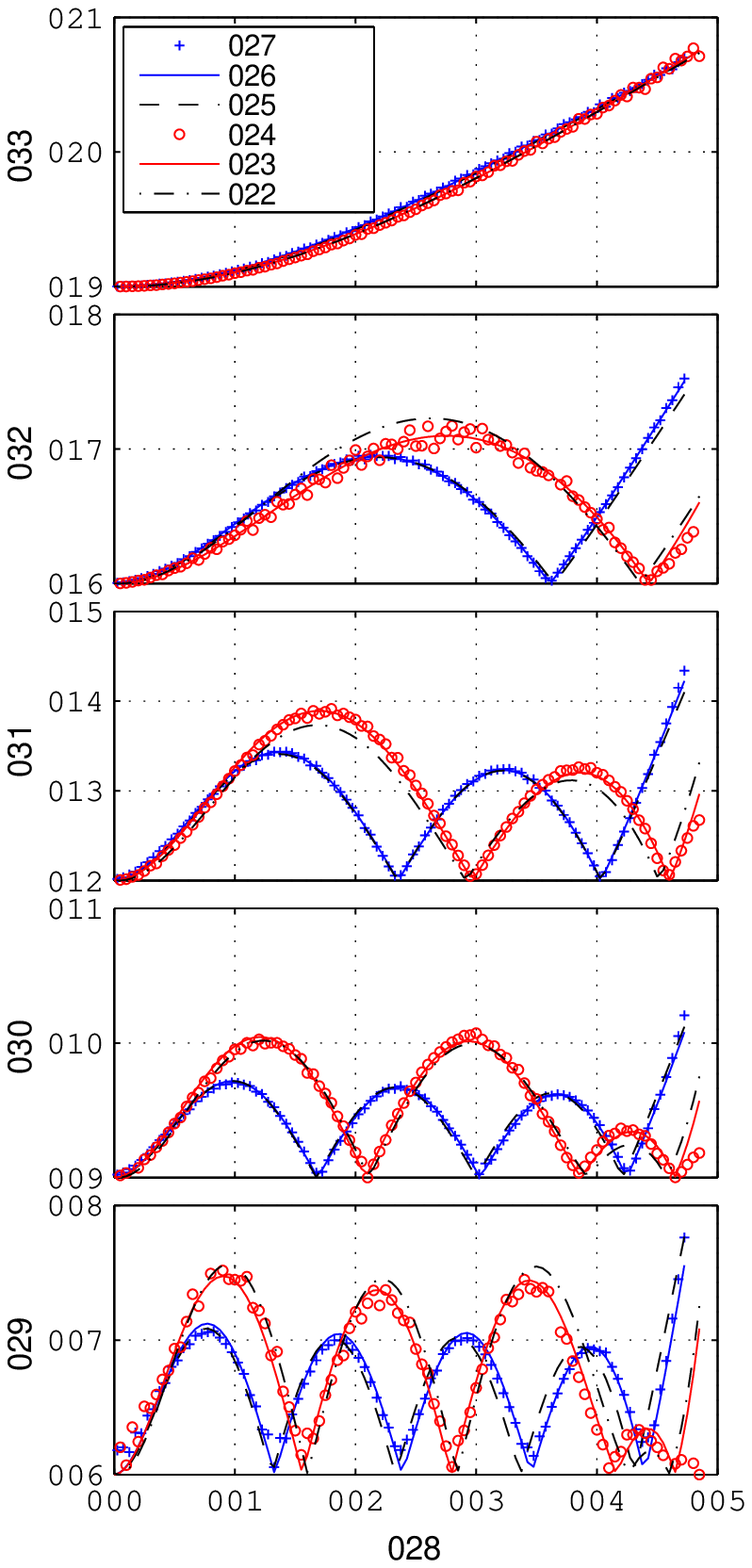}
\caption{\correction{(Colour online)} Amplitude of thermal noise for the first 5 flexural modes along the cantilever B with and without a sphere (in red and blue respectively). The markers represent the data while the lines exhibit the fits : independent fits of each mode in plain line, and simultaneous fit of all modes in dashed line. The agreement is very good for the bare and the loaded cantilever.\label{fig4}} 
\end{center}
\end{figure}

Figures~\ref{fig3} and \ref{fig4} display the rms amplitude $\sqrt{\left\langle A_n^2(x) \right\rangle}$ of the first \correction{three} resonant modes of cantilever A and the first \correction{five} modes of cantilever B respectively, with (red circles) and without \correction{(blue crosses)} the bead loading their \correction{ends}. It is worth mentioning that the maximum of the last mode is only a few \SI{}{pm} high, demonstrating the high resolution of our instrument. The behavior of the experimental data is clearly in line with the model illustrated in figure~\ref{deflexionmodek0p1Npm}: the effect of the bead is almost negligible on the first longitudinal mode, and the nodes of higher order modes are shifted towards the free end of the cantilever.

In a first step to fit the data, we use independent fits for all modes. The fitting function is equivalent to equation \ref{deflexiongenerique}. This generic function should be suitable for any cantilever clamped at its origin, whatever boundary conditions are applied at its other end. Note that $x$ is now not normalized to the length $L$ of the cantilever. We therefore use the following fitting function for $\left\langle A_n^2(x) \right\rangle$:
\begin{align}
 \psi^2(x,\bar{a},&\bar{R},\frac{\bar{\alpha}}{\bar{L}},\bar{x_0})  = \nonumber \\
 & \bar{a}^2\left(\cos\left(\bar{\alpha} \frac{x-\bar{x_0}}{\bar{L}}\right) - \cosh\left(\bar{\alpha} \frac{x-\bar{x_0}}{\bar{L}}\right)\right. \nonumber \\
 & \ \ \ \left.+ \bar{R} \left[\sin\left(\bar{\alpha} \frac{x-\bar{x_0}}{\bar{L}}\right) - \sinh \left(\bar{\alpha} \frac{x-\bar{x_0}}{\bar{L}}\right)\right]\right)^2
 \label{eq_l}
\end{align}
with $\bar{a}$, $\bar{R}$, $\bar{\alpha}/\bar{L}$ and $\bar{x_0}$ the 4 fitting parameters. As can be seen in figures~\ref{fig3} and \ref{fig4}, the result of this procedure is excellent, the model closely matching the experimental data.

The interesting output of the fit is the spatial eigenvalue $\bar{\alpha}_n/\bar{L}$ of each mode, however since $\bar{L}$ is not known precisely, only relatives values of $\bar{\alpha}_n$ can be compared to the theory. We choose to normalize the values of $\bar{\alpha}_n$ to the third mode: the presence of 2 nodes and the ``high'' amplitude constrain the fit to provide trustable values for $\bar{\alpha}_3$. The result of this procedure is displayed in tables \ref{table:alpha}. The agreement with the theoretical ratios is good for higher order modes, within a few percent. Mode 1 stands apart with a higher deviation: since no nodes are present, the fit is poorly constrained and the value of $\bar{\alpha}_1$ is not trustable.

\begin{table*}[htdp]
\begin{center}
\begin{tabular}{|l||c|c||c|c|c|c|}
\hline
Bare cantilever & \multicolumn{2}{c||}{A: $\mt_{A\tip}=0.056$, $\rt_{A\tip}= 0.03$} & \multicolumn{4}{c|}{B: $\mt_{B\tip}=0$, $\rt_{B\tip}= 0$} \\
\hline
Mode number $n$ & \makebox[2.2cm][c]{1} & \makebox[2.2cm][c]{2} & \makebox[1.2cm][c]{1} & \makebox[1.2cm][c]{2} & \makebox[1.2cm][c]{4} & \makebox[1.2cm][c]{5}  \\
\hline
Measurement: $\bar{\alpha}_{n}/\bar{\alpha}_{3}$ & 0.278 & 0.599 & 0.277   & 0.602  &   1.389   & 1.760 \\
\hline
Theory: $\alpha_{n}(\mt_{\tip},\rt_{\tip})/\alpha_{3}(\mt_{\tip},\rt_{\tip})$ & 0.236 & 0.595 & 0.239  &  0.598 &  1.400   & 1.800 \\
\hfill \emph{\small disagreement} & -18\% & -0.7\% & -15\%  &     -0.7\%  &  0.8\%  &   2.2\% \\
\hline
\hline
Loaded cantilever & \multicolumn{2}{c||}{A: $\mt_{A}=0.35$, $\rt_{A}= 0.03$} & \multicolumn{4}{c|}{B: $\mt_{B}=1.18$, $\rt_{B}= 0.06$} \\
\hline
Mode number $n$ &1 & 2 &1 & 2 & 4 & 5 \\
\hline
Measurement: $\bar{\alpha}_{n}/\bar{\alpha}_{3}$ & 0.251 & 0.580 & 0.191  &  0.561 &   1.389   & 1.865 \\
\hline
Theory: $\alpha_{n}(\mt,\rt)/\alpha_{3}(\mt,\rt)$ & 0.210 & 0.580 & 0.193   & 0.616  &    1.356  &  1.801 \\
\hfill \emph{\small disagreement} & -19\% & 0.0\% & -1.1\% &     8.8\%  &   -2.7\%  &    -3.5\% \\
\hline
\end{tabular}
\end{center}
\caption{Eigenvalues normalized to the third mode for both cantilevers, bare and loaded with the bead: we compare the output of generic fits with equation \ref{eq_l} and the values expected from the model. The agreement is quite good for every modes but the first one, where the fit is not sufficiently constrained to converge to accurate estimations.}
\label{table:alpha}
\end{table*}

\subsubsection{Stiffness determination} \label{subsubsection:stiffness}

If the generic independent fits are interesting to compare the eigenvalues and the shape of the eigenmodes with the theory, it is not possible to compare the amplitude of the modes to the expectation from thermal noise excitation. Indeed, the normalization of the mode depends on the boundary conditions (kinetic energy of the bead), and cannot be guessed a priori. To go further, we therefore perform a \emph{simultaneous fit of all modes}, imposing the values of $\alpha_n$~\cite{Paolino-JAP-2009}. The fitting function for $\left\langle A_n^2(x) \right\rangle$  \correction{is} now 
\begin{equation}
 \varphi_n^2(x,\bar{k},\bar{L} ,\bar{x_0})= \frac{3}{\alpha_n^4(\mt,\rt)} \frac{k_B T}{\bar{k}}\phi_n^2\left(\frac{x-\bar{x_0}}{\bar{L}},\mt,\rt\right)
 \label{eq:varphin}
\end{equation}
where $\phi_n(x,\mt,\rt)$  \correction{is} the normal mode defined and normalized in section \ref{section:orthogonality} (we write here explicitly the dependence in $\mt$ and $\rt$ to underline that those $\phi_n$ depend on the presence of the bead). The fitting parameters are $\bar{k}$, $\bar{L}$ and $\bar{x_0}$, the values of $\mt$ and $\rt$ are set to the estimation of section \ref{section:freq2}.

We shall thus perform a single simultaneous fit of all modes with those 3 free parameters on each cantilever, bare or loaded with a bead. We tried several weighting of the modes to compensate the decreasing amplitude of higher order modes driven by thermal noise: the function $\epsilon_j$ to minimize during the fit is defined as
\begin{equation}
 \epsilon_j(\bar{k},\bar{L} ,\bar{x_0})= \sum_n \alpha_n^j \int_0^L dx \left|\left\langle A_n^2(x)\right\rangle - \varphi_n^2(x,\bar{k},\bar{L} ,\bar{x_0})\right|^2
\end{equation}
where $j$ is a weighting parameter: $j=0$ corresponds to natural weighting (mode 1 dominant), and increasing $j$ weights more and more the higher order modes (``flat'' weights for $j=8$). We estimate the best fitting parameters $\bar{k}$, $\bar{L}$ and $\bar{x_0}$ for $j=0$ to $12$, to test the robustness of the simultaneous fitting procedure. 
The dashed lines in figures~\ref{fig3} and \ref{fig4} represent the result of this fitting process. Though not as perfect as independent fits, the results are in good agreement for all modes of each cantilever, with and without the bead.

For the bare cantilever A, the best fit values are $\bar{k}=\SI{0.321\pm0.008}{N/m}$ and $\bar{L}=\SI{497\pm1}{\micro m}$. The uncertainties correspond to the standard deviation in the full range of weighting parameter $j$, their low values demonstrating the robustness of the fit. For the loaded cantilever A, the best fit values are $\bar{k}=\SI{0.339\pm0.01}{N/m}$ and $\bar{L}=\SI{485\pm7}{\micro m}$. The stiffness of the cantilever experiences a small increase ($6\%$) after the gluing of the bead. This is not surprising since the glue increases the rigidity of the end of the cantilever, shortening its effective length by about $\SI{10}{\micro m}$ according to the fit. This $2\%$ decrease in length translates into a $6\%$ rise of the stiffness ($k$ scales as $1/L^3$), in agreement with our estimation.

For the bare cantilever B, the best fit values are $\bar{k}=\SI{0.151\pm0.002}{N/m}$ and $\bar{L}=\SI{463\pm4}{\micro m}$. Again, the dispersion of estimated parameters is very low, hinting at the robustness of the model and fitting procedure. The value of $\bar{L}$ is quite small with respect to the manufacturer specifications, however this effective length takes into account the triangular end of the cantilever. For the loaded cantilever B the best fit values are $\bar{k}=\SI{0.169\pm0.01}{N/m}$ and $\bar{L}=\SI{448\pm5}{\micro m}$. Again, this $12\%$ increase of the static spring constant goes in the expected direction, and is coherent with a reduction of the effective length of the cantilever by $\SI{15}{\micro m}$ (which should translate into a $10\%$ increase in $k$).

\section{Conclusion\label{section:ccl}}

Our work demonstrates that even a glued bead as large as $10\%$ of the length of a soft cantilever ($k\sim\SI{0.1}{N/m}$) modifies only slightly its first flexural mode and its static stiffness. A simultaneous fit of the thermal modes lead to a small decrease in the effective length and a small increase in the effective static stiffness of a cantilever upon gluing the bead. However, since these variations respect the $1/L^3$ scaling of $k$, the stiffness at the geometrical free end of the cantilever appears to be the same with and without a bead loading. Thus, the classic method to determine the spring constant by measuring the thermal spectrum of the \correction{first} flexural mode at the free end of the cantilever can still be used even if the microlever is functionalized. In fact, we have shown that the \correction{thermal calibration based on the first mode only} gets even closer to the static stiffness when the size of the bead increases, since the first mode gathers a higher fraction of the thermal energy at the free end. We have also shown that the mass model considering the load as a modification of the boundary conditions at the free end of a beam is a good approximation and  \correction{fits} well \correction{with} all our results. On the way, we have introduced a proper normalization method of the resonant modes, an unavoidable step to compute the thermal noise amplitude of each mode. Once the stiffness of the colloidal probe is determined through the classic thermal noise calibration, one will make sure to take into account other corrections due to the tip geometry, as demonstrated \correction{by Edwards and coauthors}~\cite{Edwards-2008}.

One of the main difficulties in testing the models is to estimate correctly the mass and gyration radius of the glued bead.  We have seen that it can be important to consider the initial loading due to the AFM tip and the effective length shortening in order to get consistent results: an error of $30\%$ on $\mt$ could have been made by only considering the frequency shift due to the bead, if one refers to a tipless cantilever. The ratio between resonant frequencies of a cantilever offers an interesting way to estimate the load properties (without any prior knowledge  \correction{about} the cantilever, like its unloaded properties), by comparison to tabulated values of $\alpha_n(\mt,\rt)$. We provide a set of such values in appendix~\ref{appendix:plot-alpha}, for the first 5 modes, $0 \le \mt \le 2$ and $0 \le \rt \le 0.1$ 

\begin{figure}
\begin{center}
\include{4Q}\includegraphics{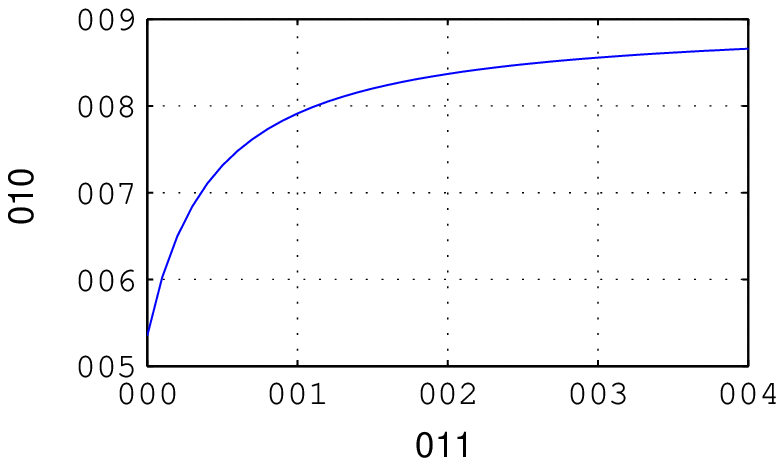}
\caption{Multiplicative correction factor to apply to the thermal noise measurement of the first resonant mode with an angular deflection measurement technique to compute the cantilever stiffness: the higher the added mass, the smaller the correction.\label{fig:4Q}} 
\end{center}
\end{figure}

As a final remark, let us study how our findings apply to the common AFM detection scheme. Indeed, our differential interferometer allows us to measure the actual deflection in any point of the cantilever, whereas most  \correction{AFMs} use an optical angular deflection measurement. The latter technique needs to be calibrated to infer from the 4  \correction{quadrant} photodiode output the true deflection of the probe. This step is usually done by a rigid contact between the probe and a hard surface, the calibrated displacement of the sample providing a benchmark to measure the sensitivity (in $\SI{}{nm/V}$). This sensitivity is thus valid for a static deformation only, and a mode dependent correction factor must be applied to estimate the actual deflection of a resonant mode~\cite{2005butt}. This multiplicative factor is 0.817 when one wants to use the thermal noise measurement of  \correction{the first mode} to calibrate the spring constant of a classic cantilever. In the current framework, we can easily compute how this correction factor depends on the normalized bead mass $\mt$ by comparing the slope of the first eigenmode to that of a static deflection with the same deflection at its free end. We plot the result in figure~\ref{fig:4Q} (computed with $\rt=0$): the correction rapidly vanishes as $\mt$ increases. Indeed, we have seen that the effect of the added mass is to decrease the eigenvalue $\alpha_1$ (see figure~\ref{alphan}), thus equation \ref{ED} tends to $z_1^{(4)}=0$ when $\mt$ increases. This last equation is that describing the static deflection, thus the first normal mode tends to the static deflection when $\mt$ increases. The effect of the added mass is thus twice \correction{beneficial} for the angular measurement: both the sensitivity of the sensor tends to be more accurate (figure~\ref{fig:4Q}), and the first mode gathers most of the thermal noise at the cantilever free end (figure~\ref{energie}). More generally, an accurate coefficient can be extracted from our analysis for any $\mt$, and applied to the thermal noise calibration of AFM colloidal probes in any commercial  \correction{device}\footnote{Integrated commercial AFM software may include calibration coefficients corresponding to a bare cantilever (most probably tipless), that should be taken into account for precise calibration of photodiode sensitivity and cantilever stiffness.}.

\begin{acknowledgments}
We are grateful to Cl\'emence Devailly for valuable assistance during the experiments, and Sergio Ciliberto for our fruitful scientific interactions. This work has been funded by ERC project Outeflucop.
\end{acknowledgments}

\appendix

\section{Spatial eigenvalues $\alpha_n(\mt,\rt)$} \label{appendix:plot-alpha}

\begin{figure}[tb]
\begin{center}
\include{alphan}\includegraphics{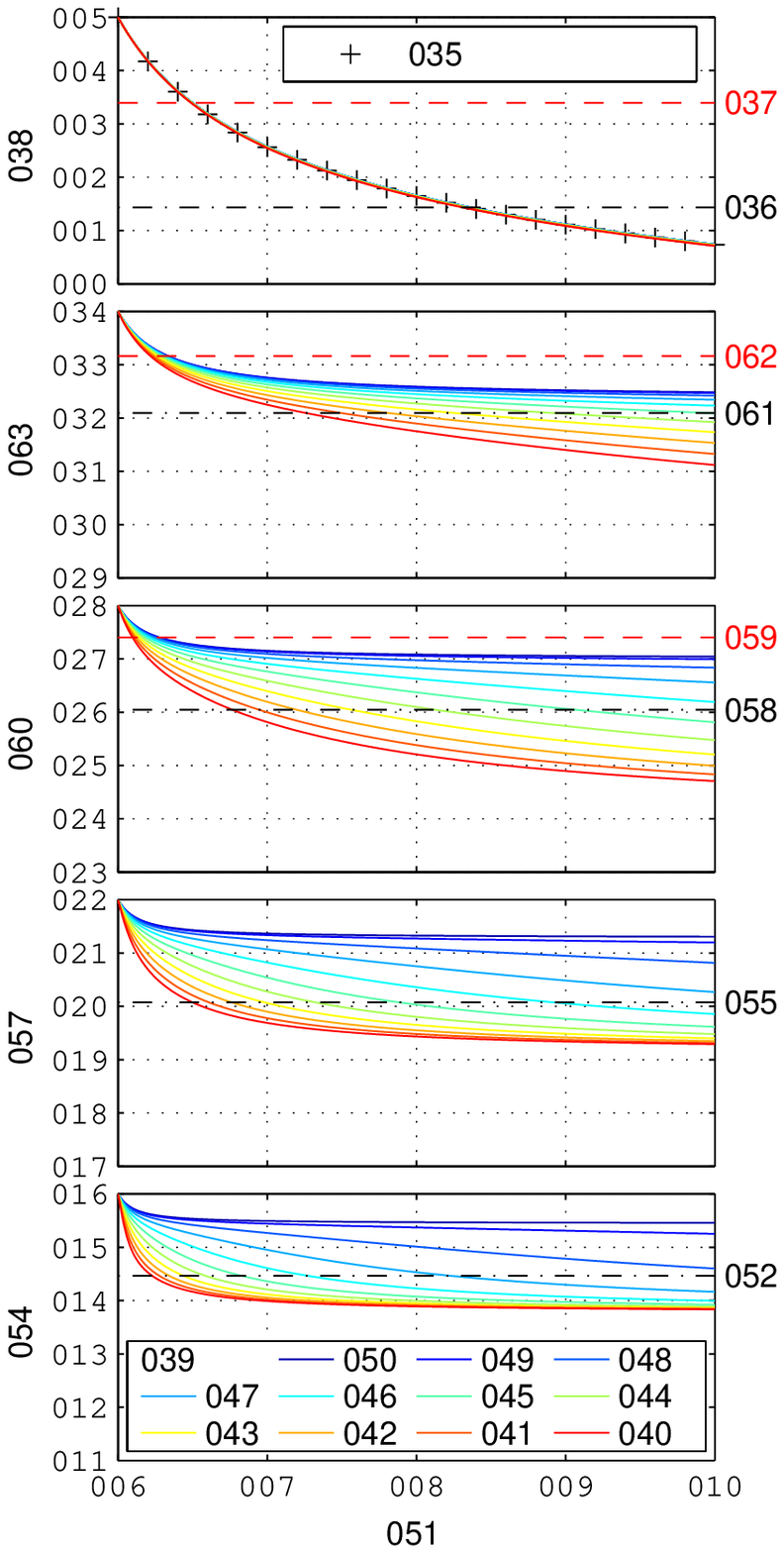}
\caption{\correction{(Colour online)} Spatial eigenvalues $\alpha_n(\mt,\rt)$ numerically computed for modes 1 to 5 as a function of $\mt$ and $\rt$ \correction{($\rt$ increases from 0 to 0.1 from top to bottom curve in each plot)}, normalized to their value at $\mt=0$: . The values of these ratios for cantilevers A and B, estimated by the frequency shift due to the addition of the bead and supposing an effective length decrease of $2\%$, are plotted as a labelled horizontal line. Mode 1 is almost independent in the value of $\rt$, and can thus be used to estimate $\mt$. This curve is very close to the prediction given in the Cleveland method~\cite{Cleveland-1993}.}
\label{alphan}
\end{center}
\end{figure}

In this appendix, we plot in figure \ref{alphan} the spatial eigenvalues $\alpha_n(\mt,\rt)$ numerically computed for modes 1 to 5 as a function of $\mt$ and $\rt$, normalized to their value at $\mt=\rt=0$. The curve for $\alpha_1$ is very close to the prediction given in the Cleveland method~\cite{Cleveland-1993}, linking the frequency shift to the added mass. This figure can be used to estimate $\mt$ and $\rt$ from the value of the frequency shift of various modes, if the initial situation corresponds to a rectangular tipless cantilever and the potential effective length decrease after gluing the particle is known.

\begin{figure}[tb]
\begin{center}
\include{alphar}\includegraphics{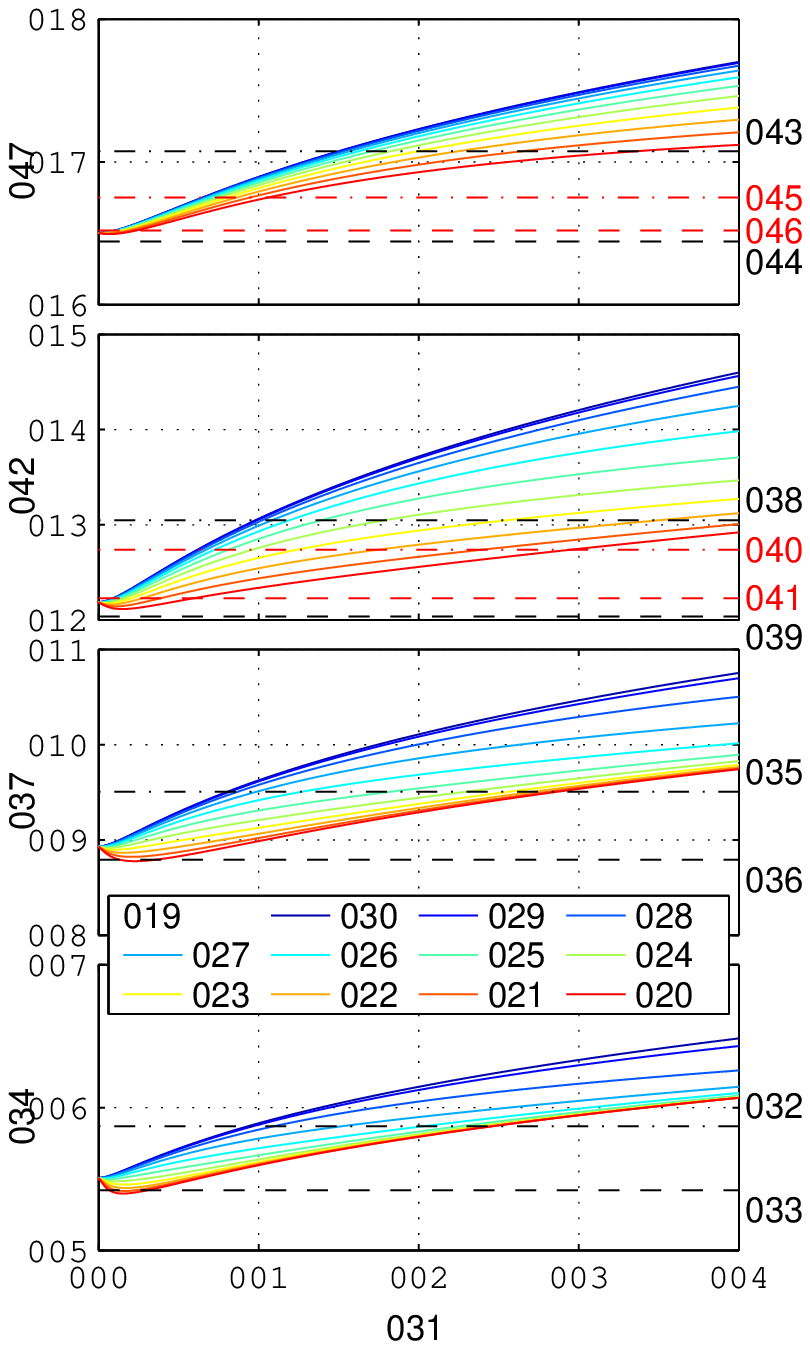}
\caption{\correction{(Colour online)} Spatial eigenvalues $\alpha_n(\mt,\rt)$ numerically computed for modes 1 to 5 as a function of $\mt$ and $\rt$ \correction{($\rt$ increases from 0 to 0.1 from top to bottom curve in each plot)}, normalized to the value of mode 1. The values of these ratios for cantilevers A and B, estimated by the frequency ratios between modes, are plotted as a labelled horizontal line for unloaded (index~$\wo$) and loaded (index$\w$) cantilevers. The model is not suited to the description of bare cantilever B (\Bwo), but allows the estimation of $\mt$ and $\rt$ for the 3 other situations.}
\label{alphar}
\end{center}
\end{figure}

In figure \ref{alphar}, we plot the same computed eigenvalues $\alpha_n(\mt,\rt)$, but normalized to the value of the first mode for the same added mass $\alpha_1(\mt,\rt)$. When one has no prior knowledge of the unloaded resonant frequencies, the ratio of the resonant frequencies between modes can be used to estimate $\mt$ and $\rt$. 

\end{document}

%\preprint{}

\title{Functionalized AFM probes for force spectroscopy: eigenmodes shape and stiffness calibration through thermal noise measurements} %Title of paper



\author{Justine Laurent, Audrey Steinberger and Ludovic Bellon}
\email[]{Ludovic.Bellon@ens-lyon.fr}
%\thanks{}
\affiliation{Universit\'e de Lyon, Laboratoire de Physique\\ \'Ecole Normale Sup\'erieure de Lyon, CNRS\\ 46 all\'ee d'Italie, FR 69007, Lyon, France}

\date{\today}

\begin{abstract}
\vspace{1cm}
\centerline{\large \bf Supplementary data}
\vspace{0.5cm}

Tables \ref{table:alpha1} to \ref{table:alpha5} report numerical values of $\alpha_n(\mt,\rt)$, for the first 5 modes, $0 \le \mt \le 2$ and $0 \le \rt \le 0.1$.
\vspace{1cm}
\end{abstract}

\maketitle 

\begin{table*}[h]
\begin{center}
\begin{tabular}{|c|c||ccccccccccc|}
\multicolumn{2}{c}{} & \multicolumn{11}{c}{Mode 1: $\alpha_1(\mt,\rt)$} \\
\cline{3-13}
\multicolumn{2}{c|}{} & \multicolumn{11}{c|}{$\rt$} \\
\cline{3-13}
\multicolumn{2}{c|}{} & 0.00 & 0.01 & 0.02 & 0.03 & 0.04 & 0.05 & 0.06 & 0.07 & 0.08 & 0.09 & 0.10 \\
\cline{3-13}
\hline
& 0.00 & 1.8751 & 1.8751 & 1.8751 & 1.8751 & 1.8751 & 1.8751 & 1.8751 & 1.8751 & 1.8751 & 1.8751 & 1.8751 \\
& 0.10 & 1.7227 & 1.7227 & 1.7226 & 1.7225 & 1.7223 & 1.7221 & 1.7218 & 1.7215 & 1.7211 & 1.7207 & 1.7203 \\
& 0.20 & 1.6164 & 1.6164 & 1.6162 & 1.6161 & 1.6158 & 1.6155 & 1.6151 & 1.6146 & 1.6140 & 1.6134 & 1.6127 \\
& 0.30 & 1.5361 & 1.5361 & 1.5360 & 1.5357 & 1.5354 & 1.5350 & 1.5345 & 1.5340 & 1.5333 & 1.5325 & 1.5317 \\
& 0.40 & 1.4724 & 1.4724 & 1.4722 & 1.4720 & 1.4716 & 1.4712 & 1.4707 & 1.4700 & 1.4693 & 1.4685 & 1.4676 \\
& 0.50 & 1.4200 & 1.4199 & 1.4198 & 1.4195 & 1.4191 & 1.4187 & 1.4181 & 1.4175 & 1.4167 & 1.4158 & 1.4149 \\
& 0.60 & 1.3757 & 1.3756 & 1.3755 & 1.3752 & 1.3748 & 1.3744 & 1.3738 & 1.3731 & 1.3723 & 1.3714 & 1.3704 \\
& 0.70 & 1.3375 & 1.3374 & 1.3373 & 1.3370 & 1.3366 & 1.3362 & 1.3356 & 1.3349 & 1.3341 & 1.3332 & 1.3321 \\
& 0.80 & 1.3041 & 1.3040 & 1.3039 & 1.3036 & 1.3032 & 1.3027 & 1.3021 & 1.3014 & 1.3006 & 1.2997 & 1.2987 \\
& 0.90 & 1.2745 & 1.2744 & 1.2742 & 1.2740 & 1.2736 & 1.2731 & 1.2725 & 1.2718 & 1.2710 & 1.2700 & 1.2690 \\
$\mt$ & 1.00 & 1.2479 & 1.2479 & 1.2477 & 1.2474 & 1.2470 & 1.2465 & 1.2459 & 1.2452 & 1.2444 & 1.2435 & 1.2424 \\
& 1.10 & 1.2239 & 1.2239 & 1.2237 & 1.2234 & 1.2230 & 1.2225 & 1.2219 & 1.2212 & 1.2204 & 1.2195 & 1.2184 \\
& 1.20 & 1.2021 & 1.2020 & 1.2018 & 1.2016 & 1.2012 & 1.2007 & 1.2001 & 1.1994 & 1.1986 & 1.1976 & 1.1966 \\
& 1.30 & 1.1820 & 1.1820 & 1.1818 & 1.1815 & 1.1812 & 1.1807 & 1.1801 & 1.1794 & 1.1785 & 1.1776 & 1.1766 \\
& 1.40 & 1.1636 & 1.1635 & 1.1633 & 1.1631 & 1.1627 & 1.1622 & 1.1616 & 1.1609 & 1.1601 & 1.1591 & 1.1581 \\
& 1.50 & 1.1464 & 1.1464 & 1.1462 & 1.1460 & 1.1456 & 1.1451 & 1.1445 & 1.1438 & 1.1430 & 1.1421 & 1.1410 \\
& 1.60 & 1.1305 & 1.1305 & 1.1303 & 1.1300 & 1.1297 & 1.1292 & 1.1286 & 1.1279 & 1.1271 & 1.1261 & 1.1251 \\
& 1.70 & 1.1156 & 1.1156 & 1.1154 & 1.1152 & 1.1148 & 1.1143 & 1.1137 & 1.1130 & 1.1122 & 1.1113 & 1.1103 \\
& 1.80 & 1.1017 & 1.1016 & 1.1015 & 1.1012 & 1.1008 & 1.1003 & 1.0998 & 1.0991 & 1.0983 & 1.0974 & 1.0963 \\
& 1.90 & 1.0886 & 1.0885 & 1.0884 & 1.0881 & 1.0877 & 1.0872 & 1.0866 & 1.0860 & 1.0852 & 1.0843 & 1.0833 \\
& 2.00 & 1.0762 & 1.0761 & 1.0760 & 1.0757 & 1.0753 & 1.0749 & 1.0743 & 1.0736 & 1.0728 & 1.0719 & 1.0709 \\
\hline
\end{tabular}
\end{center}
\caption{$\alpha_1(\mt,\rt)$: table of eigenvalues of mode 1 for $0 \le \mt \le 2$ and $0 \le \rt \le 0.1$ .}
\label{table:alpha1}
\end{table*}

\begin{table*}[htdp]
\begin{center}
\begin{tabular}{|c|c||ccccccccccc|}
\multicolumn{2}{c}{} & \multicolumn{11}{c}{Mode 2: $\alpha_2(\mt,\rt)$} \\
\cline{3-13}
\multicolumn{2}{c|}{} & \multicolumn{11}{c|}{$\rt$} \\
\cline{3-13}
\multicolumn{2}{c|}{} & 0.00 & 0.01 & 0.02 & 0.03 & 0.04 & 0.05 & 0.06 & 0.07 & 0.08 & 0.09 & 0.10 \\
\cline{3-13}
\hline
& 0.00 & 4.6941 & 4.6941 & 4.6941 & 4.6941 & 4.6941 & 4.6941 & 4.6941 & 4.6941 & 4.6941 & 4.6941 & 4.6941 \\
& 0.10 & 4.3995 & 4.3987 & 4.3964 & 4.3925 & 4.3871 & 4.3801 & 4.3715 & 4.3614 & 4.3497 & 4.3363 & 4.3215 \\
& 0.20 & 4.2671 & 4.2658 & 4.2620 & 4.2556 & 4.2467 & 4.2352 & 4.2211 & 4.2044 & 4.1852 & 4.1635 & 4.1393 \\
& 0.30 & 4.1923 & 4.1906 & 4.1856 & 4.1772 & 4.1653 & 4.1501 & 4.1315 & 4.1095 & 4.0842 & 4.0557 & 4.0241 \\
& 0.40 & 4.1444 & 4.1424 & 4.1362 & 4.1259 & 4.1114 & 4.0928 & 4.0701 & 4.0433 & 4.0125 & 3.9779 & 3.9398 \\
& 0.50 & 4.1111 & 4.1087 & 4.1015 & 4.0894 & 4.0724 & 4.0506 & 4.0239 & 3.9925 & 3.9565 & 3.9163 & 3.8722 \\
& 0.60 & 4.0867 & 4.0839 & 4.0756 & 4.0618 & 4.0424 & 4.0174 & 3.9869 & 3.9511 & 3.9102 & 3.8646 & 3.8150 \\
& 0.70 & 4.0679 & 4.0648 & 4.0555 & 4.0400 & 4.0182 & 3.9901 & 3.9559 & 3.9157 & 3.8701 & 3.8194 & 3.7647 \\
& 0.80 & 4.0531 & 4.0497 & 4.0393 & 4.0221 & 3.9980 & 3.9669 & 3.9290 & 3.8846 & 3.8343 & 3.7788 & 3.7192 \\
& 0.90 & 4.0411 & 4.0373 & 4.0260 & 4.0071 & 3.9806 & 3.9465 & 3.9049 & 3.8564 & 3.8016 & 3.7416 & 3.6775 \\
$\mt$ & 1.00 & 4.0311 & 4.0271 & 4.0148 & 3.9942 & 3.9653 & 3.9282 & 3.8831 & 3.8305 & 3.7714 & 3.7070 & 3.6386 \\
& 1.10 & 4.0228 & 4.0184 & 4.0051 & 3.9829 & 3.9517 & 3.9116 & 3.8629 & 3.8064 & 3.7431 & 3.6745 & 3.6022 \\
& 1.20 & 4.0157 & 4.0109 & 3.9967 & 3.9728 & 3.9393 & 3.8962 & 3.8440 & 3.7836 & 3.7162 & 3.6437 & 3.5677 \\
& 1.30 & 4.0096 & 4.0045 & 3.9892 & 3.9637 & 3.9278 & 3.8818 & 3.8262 & 3.7619 & 3.6907 & 3.6144 & 3.5350 \\
& 1.40 & 4.0042 & 3.9988 & 3.9826 & 3.9554 & 3.9172 & 3.8683 & 3.8092 & 3.7412 & 3.6662 & 3.5863 & 3.5038 \\
& 1.50 & 3.9995 & 3.9938 & 3.9766 & 3.9477 & 3.9073 & 3.8554 & 3.7929 & 3.7213 & 3.6427 & 3.5594 & 3.4740 \\
& 1.60 & 3.9954 & 3.9893 & 3.9711 & 3.9406 & 3.8978 & 3.8430 & 3.7772 & 3.7021 & 3.6200 & 3.5335 & 3.4454 \\
& 1.70 & 3.9916 & 3.9853 & 3.9661 & 3.9340 & 3.8889 & 3.8312 & 3.7620 & 3.6834 & 3.5980 & 3.5086 & 3.4179 \\
& 1.80 & 3.9883 & 3.9816 & 3.9615 & 3.9277 & 3.8803 & 3.8197 & 3.7473 & 3.6653 & 3.5767 & 3.4845 & 3.3915 \\
& 1.90 & 3.9853 & 3.9783 & 3.9572 & 3.9217 & 3.8720 & 3.8086 & 3.7330 & 3.6477 & 3.5560 & 3.4611 & 3.3660 \\
& 2.00 & 3.9826 & 3.9752 & 3.9531 & 3.9161 & 3.8640 & 3.7978 & 3.7190 & 3.6306 & 3.5359 & 3.4385 & 3.3414 \\
\hline
\end{tabular}
\end{center}
\caption{$\alpha_2(\mt,\rt)$: table of eigenvalues of mode 2 for $0 \le \mt \le 2$ and $0 \le \rt \le 0.1$ .}
\label{table:alpha2}
\end{table*}
 
\begin{table*}[htdp]
\begin{center}
\begin{tabular}{|c|c||ccccccccccc|}
\multicolumn{2}{c}{} & \multicolumn{11}{c}{Mode 3: $\alpha_3(\mt,\rt)$} \\
\cline{3-13}
\multicolumn{2}{c|}{} & \multicolumn{11}{c|}{$\rt$} \\
\cline{3-13}
\multicolumn{2}{c|}{} & 0.00 & 0.01 & 0.02 & 0.03 & 0.04 & 0.05 & 0.06 & 0.07 & 0.08 & 0.09 & 0.10 \\
\cline{3-13}
\hline
& 0.00 & 7.8548 & 7.8548 & 7.8548 & 7.8548 & 7.8548 & 7.8548 & 7.8548 & 7.8548 & 7.8548 & 7.8548 & 7.8548 \\
& 0.10 & 7.4511 & 7.4477 & 7.4374 & 7.4201 & 7.3956 & 7.3635 & 7.3237 & 7.2759 & 7.2202 & 7.1568 & 7.0866 \\
& 0.20 & 7.3184 & 7.3127 & 7.2956 & 7.2666 & 7.2252 & 7.1708 & 7.1032 & 7.0230 & 6.9315 & 6.8310 & 6.7248 \\
& 0.30 & 7.2537 & 7.2460 & 7.2228 & 7.1833 & 7.1265 & 7.0518 & 6.9595 & 6.8518 & 6.7324 & 6.6066 & 6.4796 \\
& 0.40 & 7.2155 & 7.2059 & 7.1769 & 7.1271 & 7.0553 & 6.9609 & 6.8456 & 6.7139 & 6.5727 & 6.4296 & 6.2910 \\
& 0.50 & 7.1903 & 7.1789 & 7.1441 & 7.0842 & 6.9974 & 6.8837 & 6.7469 & 6.5946 & 6.4368 & 6.2826 & 6.1383 \\
& 0.60 & 7.1725 & 7.1593 & 7.1187 & 7.0485 & 6.9468 & 6.8144 & 6.6577 & 6.4882 & 6.3183 & 6.1575 & 6.0114 \\
& 0.70 & 7.1593 & 7.1442 & 7.0979 & 7.0174 & 6.9007 & 6.7501 & 6.5756 & 6.3921 & 6.2137 & 6.0496 & 5.9040 \\
& 0.80 & 7.1490 & 7.1321 & 7.0800 & 6.9892 & 6.8575 & 6.6895 & 6.4991 & 6.3045 & 6.1206 & 5.9555 & 5.8121 \\
& 0.90 & 7.1408 & 7.1221 & 7.0642 & 6.9630 & 6.8164 & 6.6319 & 6.4276 & 6.2245 & 6.0373 & 5.8728 & 5.7325 \\
$\mt$ & 1.00 & 7.1341 & 7.1136 & 7.0499 & 6.9381 & 6.7769 & 6.5769 & 6.3606 & 6.1510 & 5.9623 & 5.7997 & 5.6630 \\
& 1.10 & 7.1286 & 7.1063 & 7.0367 & 6.9143 & 6.7387 & 6.5242 & 6.2977 & 6.0835 & 5.8945 & 5.7345 & 5.6019 \\
& 1.20 & 7.1239 & 7.0998 & 7.0243 & 6.8913 & 6.7015 & 6.4737 & 6.2386 & 6.0212 & 5.8331 & 5.6762 & 5.5477 \\
& 1.30 & 7.1199 & 7.0940 & 7.0126 & 6.8689 & 6.6653 & 6.4251 & 6.1830 & 5.9637 & 5.7772 & 5.6237 & 5.4994 \\
& 1.40 & 7.1164 & 7.0887 & 7.0013 & 6.8470 & 6.6299 & 6.3786 & 6.1307 & 5.9105 & 5.7262 & 5.5763 & 5.4562 \\
& 1.50 & 7.1134 & 7.0838 & 6.9905 & 6.8255 & 6.5954 & 6.3339 & 6.0814 & 5.8612 & 5.6794 & 5.5333 & 5.4172 \\
& 1.60 & 7.1108 & 7.0793 & 6.9800 & 6.8043 & 6.5616 & 6.2910 & 6.0349 & 5.8153 & 5.6364 & 5.4941 & 5.3820 \\
& 1.70 & 7.1084 & 7.0751 & 6.9697 & 6.7834 & 6.5286 & 6.2498 & 5.9910 & 5.7726 & 5.5968 & 5.4583 & 5.3500 \\
& 1.80 & 7.1063 & 7.0712 & 6.9597 & 6.7627 & 6.4964 & 6.2103 & 5.9496 & 5.7328 & 5.5602 & 5.4255 & 5.3208 \\
& 1.90 & 7.1044 & 7.0675 & 6.9499 & 6.7423 & 6.4649 & 6.1723 & 5.9104 & 5.6955 & 5.5263 & 5.3953 & 5.2941 \\
& 2.00 & 7.1027 & 7.0639 & 6.9402 & 6.7221 & 6.4341 & 6.1359 & 5.8734 & 5.6607 & 5.4949 & 5.3674 & 5.2696 \\
\hline
\end{tabular}
\end{center}
\caption{$\alpha_3(\mt,\rt)$: table of eigenvalues of mode 3 for $0 \le \mt \le 2$ and $0 \le \rt \le 0.1$ .}
\label{table:alpha3}
\end{table*}

\begin{table*}[htdp]
\begin{center}
\begin{tabular}{|c|c||ccccccccccc|}
\multicolumn{2}{c}{} & \multicolumn{11}{c}{Mode 4: $\alpha_4(\mt,\rt)$} \\
\cline{3-13}
\multicolumn{2}{c|}{} & \multicolumn{11}{c|}{$\rt$} \\
\cline{3-13}
\multicolumn{2}{c|}{} & 0.00 & 0.01 & 0.02 & 0.03 & 0.04 & 0.05 & 0.06 & 0.07 & 0.08 & 0.09 & 0.10 \\
\cline{3-13}
\hline
& 0.00 & 10.9955 & 10.9955 & 10.9955 & 10.9955 & 10.9955 & 10.9955 & 10.9955 & 10.9955 & 10.9955 & 10.9955 & 10.9955 \\
& 0.10 & 10.5218 & 10.5127 & 10.4849 & 10.4372 & 10.3676 & 10.2745 & 10.1581 & 10.0210 & 9.8698 & 9.7129 & 9.5591 \\
& 0.20 & 10.4016 & 10.3862 & 10.3387 & 10.2555 & 10.1325 & 9.9701 & 9.7771 & 9.5714 & 9.3720 & 9.1921 & 9.0377 \\
& 0.30 & 10.3480 & 10.3269 & 10.2607 & 10.1427 & 9.9681 & 9.7450 & 9.5000 & 9.2643 & 9.0586 & 8.8894 & 8.7547 \\
& 0.40 & 10.3178 & 10.2910 & 10.2061 & 10.0527 & 9.8278 & 9.5550 & 9.2796 & 9.0376 & 8.8423 & 8.6913 & 8.5764 \\
& 0.50 & 10.2984 & 10.2660 & 10.1620 & 9.9726 & 9.7003 & 9.3899 & 9.1010 & 8.8654 & 8.6860 & 8.5530 & 8.4547 \\
& 0.60 & 10.2850 & 10.2469 & 10.1235 & 9.8976 & 9.5823 & 9.2460 & 8.9550 & 8.7317 & 8.5690 & 8.4519 & 8.3669 \\
& 0.70 & 10.2751 & 10.2313 & 10.0882 & 9.8257 & 9.4728 & 9.1205 & 8.8348 & 8.6260 & 8.4788 & 8.3751 & 8.3008 \\
& 0.80 & 10.2675 & 10.2181 & 10.0547 & 9.7563 & 9.3715 & 9.0111 & 8.7349 & 8.5409 & 8.4077 & 8.3152 & 8.2494 \\
& 0.90 & 10.2615 & 10.2064 & 10.0225 & 9.6889 & 9.2779 & 8.9156 & 8.6511 & 8.4714 & 8.3503 & 8.2672 & 8.2083 \\
$\mt$ & 1.00 & 10.2566 & 10.1958 & 9.9911 & 9.6236 & 9.1919 & 8.8321 & 8.5803 & 8.4138 & 8.3033 & 8.2280 & 8.1749 \\
& 1.10 & 10.2526 & 10.1860 & 9.9602 & 9.5603 & 9.1128 & 8.7587 & 8.5199 & 8.3654 & 8.2641 & 8.1954 & 8.1471 \\
& 1.20 & 10.2492 & 10.1768 & 9.9297 & 9.4992 & 9.0403 & 8.6941 & 8.4680 & 8.3243 & 8.2309 & 8.1679 & 8.1237 \\
& 1.30 & 10.2463 & 10.1681 & 9.8994 & 9.4403 & 8.9738 & 8.6369 & 8.4229 & 8.2890 & 8.2026 & 8.1444 & 8.1037 \\
& 1.40 & 10.2438 & 10.1597 & 9.8694 & 9.3837 & 8.9128 & 8.5861 & 8.3836 & 8.2585 & 8.1781 & 8.1242 & 8.0864 \\
& 1.50 & 10.2417 & 10.1516 & 9.8396 & 9.3293 & 8.8568 & 8.5408 & 8.3490 & 8.2317 & 8.1568 & 8.1065 & 8.0713 \\
& 1.60 & 10.2398 & 10.1438 & 9.8099 & 9.2773 & 8.8054 & 8.5002 & 8.3184 & 8.2082 & 8.1380 & 8.0910 & 8.0580 \\
& 1.70 & 10.2381 & 10.1361 & 9.7804 & 9.2275 & 8.7580 & 8.4636 & 8.2912 & 8.1874 & 8.1214 & 8.0772 & 8.0463 \\
& 1.80 & 10.2366 & 10.1286 & 9.7511 & 9.1799 & 8.7145 & 8.4307 & 8.2668 & 8.1687 & 8.1066 & 8.0650 & 8.0358 \\
& 1.90 & 10.2352 & 10.1212 & 9.7220 & 9.1345 & 8.6743 & 8.4008 & 8.2449 & 8.1521 & 8.0933 & 8.0540 & 8.0264 \\
& 2.00 & 10.2340 & 10.1139 & 9.6931 & 9.0913 & 8.6371 & 8.3737 & 8.2251 & 8.1370 & 8.0813 & 8.0440 & 8.0179 \\
\hline
\end{tabular}
\end{center}
\caption{$\alpha_4(\mt,\rt)$: table of eigenvalues of mode 4 for $0 \le \mt \le 2$ and $0 \le \rt \le 0.1$ .}
\label{table:alpha4}
\end{table*}
 
\begin{table*}[htdp]
\begin{center}
\begin{tabular}{|c|c||ccccccccccc|}
\multicolumn{2}{c}{} & \multicolumn{11}{c}{Mode 5: $\alpha_5(\mt,\rt)$} \\
\cline{3-13}
\multicolumn{2}{c|}{} & \multicolumn{11}{c|}{$\rt$} \\
\cline{3-13}
\multicolumn{2}{c|}{} & 0.00 & 0.01 & 0.02 & 0.03 & 0.04 & 0.05 & 0.06 & 0.07 & 0.08 & 0.09 & 0.10 \\
\cline{3-13}
\hline
& 0.00 & 14.1372 & 14.1372 & 14.1372 & 14.1372 & 14.1372 & 14.1372 & 14.1372 & 14.1372 & 14.1372 & 14.1372 & 14.1372 \\
& 0.10 & 13.6142 & 13.5953 & 13.5364 & 13.4314 & 13.2743 & 13.0670 & 12.8284 & 12.5890 & 12.3745 & 12.1966 & 12.0553 \\
& 0.20 & 13.5067 & 13.4742 & 13.3700 & 13.1781 & 12.8973 & 12.5727 & 12.2730 & 12.0358 & 11.8619 & 11.7373 & 11.6475 \\
& 0.30 & 13.4615 & 13.4160 & 13.2659 & 12.9859 & 12.6053 & 12.2337 & 11.9480 & 11.7515 & 11.6196 & 11.5296 & 11.4665 \\
& 0.40 & 13.4367 & 13.3782 & 13.1803 & 12.8138 & 12.3660 & 11.9938 & 11.7428 & 11.5833 & 11.4805 & 11.4116 & 11.3637 \\
& 0.50 & 13.4210 & 13.3494 & 13.1019 & 12.6552 & 12.1717 & 11.8212 & 11.6053 & 11.4741 & 11.3911 & 11.3359 & 11.2975 \\
& 0.60 & 13.4102 & 13.3253 & 13.0267 & 12.5097 & 12.0149 & 11.6941 & 11.5082 & 11.3981 & 11.3291 & 11.2833 & 11.2514 \\
& 0.70 & 13.4023 & 13.3040 & 12.9532 & 12.3778 & 11.8882 & 11.5980 & 11.4367 & 11.3425 & 11.2837 & 11.2447 & 11.2175 \\
& 0.80 & 13.3963 & 13.2843 & 12.8808 & 12.2594 & 11.7854 & 11.5236 & 11.3821 & 11.3002 & 11.2491 & 11.2151 & 11.1914 \\
& 0.90 & 13.3916 & 13.2658 & 12.8096 & 12.1539 & 11.7012 & 11.4646 & 11.3392 & 11.2669 & 11.2218 & 11.1918 & 11.1709 \\
$\mt$ & 1.00 & 13.3878 & 13.2479 & 12.7398 & 12.0603 & 11.6314 & 11.4169 & 11.3047 & 11.2401 & 11.1998 & 11.1730 & 11.1542 \\
& 1.10 & 13.3846 & 13.2305 & 12.6715 & 11.9774 & 11.5730 & 11.3777 & 11.2764 & 11.2181 & 11.1817 & 11.1574 & 11.1404 \\
& 1.20 & 13.3820 & 13.2134 & 12.6051 & 11.9040 & 11.5237 & 11.3449 & 11.2527 & 11.1997 & 11.1665 & 11.1444 & 11.1288 \\
& 1.30 & 13.3797 & 13.1966 & 12.5408 & 11.8390 & 11.4816 & 11.3172 & 11.2327 & 11.1841 & 11.1536 & 11.1333 & 11.1190 \\
& 1.40 & 13.3778 & 13.1798 & 12.4788 & 11.7811 & 11.4454 & 11.2934 & 11.2155 & 11.1707 & 11.1425 & 11.1237 & 11.1104 \\
& 1.50 & 13.3761 & 13.1632 & 12.4193 & 11.7296 & 11.4139 & 11.2728 & 11.2007 & 11.1590 & 11.1329 & 11.1154 & 11.1030 \\
& 1.60 & 13.3746 & 13.1465 & 12.3622 & 11.6836 & 11.3863 & 11.2549 & 11.1876 & 11.1489 & 11.1244 & 11.1081 & 11.0965 \\
& 1.70 & 13.3733 & 13.1299 & 12.3078 & 11.6424 & 11.3620 & 11.2391 & 11.1762 & 11.1398 & 11.1170 & 11.1016 & 11.0907 \\
& 1.80 & 13.3721 & 13.1132 & 12.2558 & 11.6052 & 11.3404 & 11.2251 & 11.1660 & 11.1318 & 11.1103 & 11.0958 & 11.0856 \\
& 1.90 & 13.3711 & 13.0966 & 12.2065 & 11.5718 & 11.3212 & 11.2125 & 11.1569 & 11.1247 & 11.1043 & 11.0906 & 11.0810 \\
& 2.00 & 13.3701 & 13.0798 & 12.1596 & 11.5414 & 11.3039 & 11.2013 & 11.1487 & 11.1182 & 11.0989 & 11.0860 & 11.0768 \\
\hline
\end{tabular}
\end{center}
\caption{$\alpha_5(\mt,\rt)$: table of eigenvalues of mode 5 for $0 \le \mt \le 2$ and $0 \le \rt \le 0.1$ .}
\label{table:alpha5}
\end{table*}